\documentclass[suppldata]{interact}
\usepackage[utf8x]{inputenc}
\usepackage[english]{babel}
\usepackage{ucs}
\usepackage[authoryear,round]{natbib}
\usepackage{amsmath,amsfonts,bm,amssymb,latexsym,amsthm}
\usepackage{graphicx, graphics}
\usepackage{multirow}
\usepackage{makecell}
\usepackage{rotating}
\usepackage{microtype}
\usepackage[none]{hyphenat}
\usepackage{setspace} 
\usepackage{array}
\usepackage{amsfonts}
\usepackage[bottom]{footmisc}
\usepackage{color}
\usepackage[table]{xcolor}
\usepackage{url}
\usepackage{float}
\usepackage{rotating}
\usepackage{mathtools}
\usepackage{adjustbox}
\usepackage[table]{xcolor}
\usepackage{textcomp}
\usepackage{placeins}
\usepackage{multicol}
\usepackage{multirow}
\usepackage{enumerate}
\usepackage{longtable}
\usepackage{morefloats}
\usepackage{booktabs}
\usepackage{multirow}
\usepackage[para,online,flushleft]{threeparttable}
\usepackage[justification=justified,font=small]{caption}
\usepackage{pifont}
\usepackage[show]{chato-notes}
\usepackage{subcaption} 
\usepackage{hyperref}
\usepackage{cleveref}
\usepackage{svg}

\usepackage{footnote}
\usepackage[flushleft]{threeparttable}

\newcolumntype{C}[1]{>{\centering\let\newline\\\arraybackslash\hspace{0pt}}m{#1}}

\hypersetup{colorlinks, breaklinks, linkcolor=[rgb]{0,0.3,1}, citecolor=[rgb]{0,0.3,1}, urlcolor=[rgb]{0,0.3,1} }
\usepackage{longtable}

\title{Measuring Corporate Digital Divide with web scraping: Evidence from Italy}

\author{
\name{Leonardo Mazzoni\textsuperscript{a,b}\thanks{Corresponding Author: Leonardo Mazzoni.  Email: leonardo.mazzoni@imtlucca.it}, Fabio Pinelli\textsuperscript{b} and Massimo Riccaboni\textsuperscript{b}}
\affil{\textsuperscript{a} University of Padua - Department of Economics and Management, Padua (Italy);\\ \textsuperscript{b}IMT School for Advanced Studies, Lucca (Italy)}
}

\begin{document}

\maketitle

\begin{abstract}
With the increasing pervasiveness of ICTs in the fabric of economic activities, the corporate digital divide has emerged as a new crucial topic to evaluate the IT competencies and the digital gap between firms and territories. Given the scarcity of available granular data to measure the phenomenon, most studies have used survey data. To bridge the empirical gap, we scrape the website homepage of 182\,705 Italian firms, extracting ten features related to their digital footprint  characteristics to develop a new corporate digital assessment index. Our results highlight a significant digital divide across dimensions, sectors and
geographical locations of Italian firms, opening up new perspectives on monitoring and
near-real-time data-driven analysis. 

\end{abstract}
\medskip 


\noindent \textbf{Acknowledgements}: The authors would like to thank the following projects for the financial support and comments received: Artes 4.0 -Advanced Robotics and enabling digital Technologies \& Systems; "Rinascita dei Borghi" funded by Eurispes (Institute for Political, Economic and Social Studies) and the Italian Ministry of Economy and Finance; "Borghi, paesi, aree interne: infrastrutture, sostenibilità e qualità della vita" funded by the Italian Ministry of University and Research.

\pagebreak

\section[sec:intro]{Introduction}

Digital transformation has recently emerged as a driving force able to forge the strategic orientation of firms to grow and innovate by means of digital technologies and the relative capabilities built on them \citep{Blankaetal2020, Verhoefetal2021, Volberdaetal2021}. With the increasing pervasiveness of ICTs in the fabric of economic activities \citep{Antonelli2003,Baskervilleetal2020}, a heterogeneous response by individuals, firms and institutions has occurred, translating into different rates of adoption of, and proficiency with, digital tools. This phenomenon has been generally analyzed under the umbrella of the Digital Divide, a concept able to represent the gap in accessing IT infrastructure \citep{FinkandKenny2003}. Afterwards, the capillary diffusion of ICTs contributed to extend the Digital Divide definition also in its usage, involving dedicated human resources and digital market providers \citep{CorrocherandOrdanini2002, Kyriakidouetal2011}. 
In this respect, the literature on Digital Divide has not developed a homogenous corpus of analysis. New forms are rapidly emerging with growing attention to its specific aspects for industries, firms, and territorial levels \citep{Ellingeretal2003, Shakinaetal2021, lythreatis2022, Thoniparaetal2022}.
Particularly, the corporate Digital Divide is a crucial emerging topic in the management literature for the consequences brought by the Industry 4.0 paradigm on the competencies to develop \citep{Shakinaetal2021}. As yet, the literature on the Digital Divide has remained relatively silent on the mechanisms occurring at the firm level, with very few studies addressing this specific subject \citep{lythreatis2022}. This can be attributed to the fact that the Corporate Digital Divide is still difficult to observe for the lack of accounting metrics able to provide information on IT investment or the implementation of information systems updating in the cognitive architecture of the firm \citep{Vehovaretal2006, Tambeetal2020}.

A possible way out from data shortage on the digital behavior of firms came from the analysis of corporate websites \citep{BlazquezandDomenech2018}. Accordingly, corporate websites are the ``digital footprint'' of organizations and part of new codified knowledge, which is increasingly becoming accessible for researchers and analysts to study the performance of firms in complementary/additional ways, with respect to the more traditional data sources \citep{Goketal2015, Blazquezetal2018, KinneandResch2018}. This is because websites represent the self-expression of strategic information to external stakeholders: the products/services commercialized, delivery modes, mission and vision, the internal competencies, the relationships with other companies and universities, research activities, their location and facilities \citep{Youtieetal2012, Goketal2015, Lietal2018, Saridakisetal2018, PukelisandStanciauskas2019}. Moreover, new decision-making procedures, cost structures, organizational routines and digital operations have been consequently introduced \citep{TeeceandLiden2017, Verhoefetal2021}.
This makes websites, especially corporate ones, an essential open data source not only to capture the visibility and reputation of the firm but also to study the broader digital competencies beyond them \citep{auger2005}. Despite a firm may rely on software houses or external IT consultants to build their websites, the specific  technical features that characterize them imply indirect agency to transform digital objects by means of  ``sensing capabilities'' on the intrinsic value of digital technologies \citep{FaulknerandRunde2009}.
Accordingly, some studies have exploited the characteristics of websites as wider signals of the digital awareness of firms \citep{wellsetal2011, abeysekera2019}.

 Thanks to the recent evolution of web-scraping techniques  \citep{Aroraetal2016, AxenbeckBreithaupt2019, Aroraetal2020}, our aim is to leverage information extracted from the corporate website to study the corporate Digital Divide at a large scale, considering different firms' characteristics (e.g., dimension, industry, age, and geographical context). 
More concretely, we scraped the websites of 182\,705 Italian firms in the period 2020-2021, extracting ten features related to the technical libraries, performances, security level, speed, links and social media. Next, we analyzed corporate website features in combination with the relative corporate information. Instead of considering the contents of websites, we focused on the most  ``objective'' IT features available, following the research line aimed to exploit new IT technologies as new economic-related proxies of digital capabilities \citep{george2016, brynjolfssonetal2021}.

We analyze the Italian case because of the well-known sharp socio-economic disparities between northern and southern regions \citep{Daniele2021}. Moreover, Italy is a unique case in Europe of an industrialized country lagging behind other EU countries in terms of digital readiness.\footnote{See the results of the Digital Economy Society index available at:
https://digital-strategy.ec.europa.eu/en/policies/desi.}

Our results highlight a significant corporate Digital Divide across firms' attributes such as dimension, sector and age, and territorial characteristics where the firm is located, opening up new perspectives of monitoring and near-real-time data-driven analysis. Controlling for the impact of wide band our results still hold, paving the way for further empirical investigations.

Previous research has spotted a shortcoming of studies in the analysis of corporate digital behaviour with large website samples \cite[e.g.,][]{lythreatis2022}.
Our study contributes to fill this gap, nuancing the current understanding of firms' Digitial Divide, by exploiting big open-source data directly extracted by corporate websites.
The extracted features described a multifaceted phenomenon. Interestingly as evidenced by low correlation among different elements, the digital-related variables capture specific capabilities and justify this explorative analysis. However, in order to ensure comparability among firms (and of territories), we propose an aggregation of the ten features, interpreting and categorizing them according to a theoretical building process based on the digital space(s) of the firm: technical capabilities, internal organizations,external stakeholder engagement, and digital culture.


The paper is organized as follows. Section 2 reviews the literature on the economic and Digital Divide and the role of websites in measuring digital footprint. Section 3 describes the data collection process and the methodology adopted. Section 4 displays the results of the empirical investigation of the Digital Divide phenomenon. Section 5 discusses the findings, underlining the limitations of this work and concluding with final remarks.
\section{Literature review}

\subsection{The Digital Divide notion: a corporate perspective}

The notion of the Digital Divide was initially coined as the different rates of adoption of ICTs of individuals and households \citep{FinkandKenny2003, Vehovaretal2006, Kyriakidouetal2011}. Then the massive diffusion of the internet has shifted the attention beyond the simple rate of adoption, reaching a further layer related to the usage of ICTs \citep{CorrocherandOrdanini2002}. The transition from the industrial society to the information economy\citep{Castells1996} and the recent conceptualisation of ``onlife'' societies \citep{Floridi2014} with the new role of ICTs as ``reality shapers'' \citep{Baskervilleetal2020} has magnified the relevance of this divide, as a reflection of the socio-economic gap between individuals, firms and territories. Accordingly, the meaning of the Digital Divide has become a multifaceted and more elaborated notion, including the developed skills and abilities to use technical tools \citep{FinkandKenny2003, Szeles2018, MatthesandKunkel2020}.
This has allowed elaborating more on the competitiveness drivers of the digital economy, being ICTs firmly embedded in the fabric of socio-economic systems \citep{Antonelli2003, Forman2005}. Furthermore, with the advent of the Industry 4.0 paradigm in the last decade, the concept of the digital divide has furtherly increased its significance as a metric able to reflect economic performances \citep{Shakinaetal2021}. This is particularly relevant considering the growing pervasiveness of ICTs and the complementarity between physical and key enabling digital technologies in the production and consumption of goods and services (e.g., cloud computing, artificial intelligence) \citep{Giustizieroetal2021}.
While there is a growing awareness of the digital performance of countries and regions, the notion of the Digital Divide registers very few contributions applied to the firms as units of analysis. 

As evidenced by recent contributions, digital transformation has noteworthy impacted firms’ structure, being a strategic transformation of organization and core capabilities of businesses enabled by digital technologies \citep{Volberdaetal2021}. The rapid and unceasing technological change that occurred in the last decade has challenged the status quo of firms across industries, creating gaps for different rates of digital awareness by managers and employees and diverse accumulation of digital-related skills \citep{Blankaetal2020, Shakinaetal2021}. Internal routines and relationships with customers and suppliers have been radically altered, and unsurprisingly, the term co-creation is frequently applied to refer to collaboration between the various actors in business ecosystems in the value creation path \citep{WarnerandWager2019, Liuetal2021}.
Bearing in mind this transformation, the traditional resource-based view of the firm paradigm \citep{Barney1991, Wernerfelt1984} has been profoundly impacted by digital technologies and their bundled use \citep{Giustizieroetal2021}. Accordingly, the development of new capabilities to favor business model adaptation to the new techno-economic scenario requires a digital sensing activity by the firm \citep{WarnerandWager2019}. The ubiquity of ICTs requires not only the ownership of specific resources but also the creation of specialized human resources to frame the new possibilities opened by digital affordance property, that is, the creation of endless reconfiguration by the use of the same inputs or a creative (re)combination of them \citep{Giustizieroetal2021, Liuetal2021}.
All in all, while we have rather substantial theoretical evidence that digital transformation has impacted corporates’ structure and strategic approach, we are still struggling to provide detailed and fine-grained measures at the firm level \citep{Tambeetal2020}. In other words, we cannot evaluate firms' response to the introduction of digital technologies and if they have developed adequate digital capabilities. Hence, the literature has remained relatively silent on the corporate Digital Divide level across different typologies of firms, operating in various industries, and localized in urban or peripheral contexts. 

The good news is that digital footprints left by organisations and individuals have recently become available data for empirical analyses, thanks to the diffusion in social sciences of methodologies such as web scraping \citep{Lietal2018, AxenbeckBreithaupt2019, KinneandAxenback2020, Thoniparaetal2022}. In other words, considering information reported on the internet under the lenses of digital signalling theory \citep{wellsetal2011, abeysekera2019} allows relating the produced digital artifacts (as the characteristics of a corporate website) to a set of underlying digital capabilities \citep{ageevaetal2018}.
Recent studies have exploited this caveat, investigating the relationships within the innovation ecosystem between firms, universities, and institutions \citep{Lietal2018}, the digital layer of companies and the concept of proximity \citep{Krugeretal2020}, the innovation performance of firms \citep{KinneandAxenback2020}. 
Despite this growing popularity, researches on these new data sources are still very fragmented, with many exploratory analyses and many elements that may not be applicable in all settings \citep{Hernandezetal2009}. With few exceptions \citep{KinneandAxenback2020, Krugeretal2020, Thoniparaetal2022}, most of the analyses have concentrated on small samples, without the possibility of transforming these unstructured and very heterogeneous data sources into new potential tools to investigate economic performance at the systemic level.

\subsection{The use of websites to measure the digital footprint of firms}

The wide adoption of websites among firms, including SMEs, represents an interesting potential information source to bridge the gap between the need to assess the digital performance of firms and the lack of granular indicators. This is possible for the strategic role played by websites in the information and knowledge economy.

A website is a digital means able to reduce information asymmetry between two parts, facilitating corporate operations (delivery, customer care, internationalization) \citep{Billonetal2009}. Its maintenance, use, and development imply some extra costs for the firms. This candidates it as \textit{``near-costless measure of marketing''} \citep{Thoniparaetal2022} and, more in general, as an effective proxy to capture the digital footprint of economic agents \citep{Goketal2015, BlazquezandDomenech2018, HerouxVaillancourtetal2020, KinneandAxenback2020}.
Websites are publicly available sources, manageable at any desired time, and represent a new form of codified knowledge to complement information sources on the firms' performances \citep{Blanketal2018, KinneandResch2018}. The embeddedness of websites into online environments and the introduction of e-commerce platforms have created completely new value delivery bidirectional channels \citep{Saridakisetal2018, Verhoefetal2021}. As a result, this has impacted a wide and heterogeneous set of industries.
The massive use of websites offers some advantages in data collection in comparison to traditional methodologies due to their (i) ``unobtrusiveness'', (ii) accessibility, (iii) temporal frequency, (iv) granularity, and (v) coverage  \citep{Mateosetal2001, Goketal2015, RasmussenandThimm2015, BlazquezandDomenech2018, KinneandResch2018, Lietal2018}. Unobtrusiveness derives from the capacity to directly gather the information reported without requiring the direct involvement of a firm or a set of them (e.g., surveys). This saves time collecting information and is less variable than a traditional survey. Accessibility stems from the open-access nature of websites as information sources and the reproducibility of the analysis. Websites, especially corporate websites, are updated for business reasons much more frequently than traditional information sources such as surveys (temporal frequency).
Moreover, the information reported on a website has a much more customisable degree of granularity than traditional survey methodologies (even if with increasing post-processing works). Finally, the massive analysis of websites has the potential to extend coverage to almost the entire population of companies (with the absence of non-response), overcoming the limitations of traditional collection methods based on the selection of a representative sample. This makes it possible to detect common and recurring characteristics agnostically and to identify hidden peculiarities and specificities that remain under the tip of the iceberg.
Extracting data directly from websites can present some problems regarding the reliability of the information. However, companies would face rather negative feedback from the clients and/or business partners with whom they interact if they would exaggerate or mystify the reality. Hence, the general trend within the literature is to treat this information as truthful or at least documentable \citep{PukelisandStanciauskas2019}. Notwithstanding the availability of websites, some doubts emerge on the typology of firms detectable on the web. Companies very close to the market are most likely to be included in this group (e.g., B2C). However, we have less evidence for subcontractors or intermediaries (B2B), whose activities can remain in the shadow to respect the will of their final clients (e.g., respecting industrial secrets or preventing competition) \citep{PukelisandStanciauskas2019}. Encouraging signals, in this case, come from the requesting of certificates (as ISO) by value chain leaders that are pushing subcontractors and suppliers, in general, to show them as digital ``business cards''. Still, available company websites are a convenience sample to investigate firm dynamics, and a proper procedure can be implemented to re-balance the available information across geographical areas and sectors. 
The difficulties of website data processing are related to their reliability and data mining procedures. There are still some technical problems with treating homogeneously website data, which for their nature are organized in several formats, with different dimensionality, and composed of non-textual parts (i.e., graphics and images) \citep{Goketal2015, Beaudryetal2016, HerouxVaillancourtetal2020}.
\section[sec:data]{Data Collection and Methodology}\label{sec:data}

\subsection{Data Collection}
From the AIDA database (Bureau van Dijk), we collected information about the website of Italian firms. Initially, we collected website URL information for 450\,348 firms covering the entire national territory.

We implemented in \texttt{python} our crawler script using a combination of different libraries: \texttt{requests}\footnote{docs.python-requests.org}, \texttt{scrapy}\footnote{https://scrapy.org/}, \texttt{beautifulsoup}\footnote{https://www.crummy.com/software/BeautifulSoup/bs4/doc/}, \texttt{trafilatura}\footnote{https://trafilatura.readthedocs.io/en/latest/}, \texttt{builtwith}\footnote{https://pypi.org/project/builtwith/} and finally we also rely on the results of Google Lighthouse\footnote{https://developers.google.com/web/tools/lighthouse}. Next, we stored the results of the crawling activities on a \texttt{MongoDB}\footnote{ https://www.mongodb.com/} instance, a document-oriented NOSQL database that provided us the right flexibility to store and update the gathered information. 
Each firm's features were stored in a single \texttt{MongoDB} document. 

The crawling activity has been carried out from January 2021 to March 2021, with a further update in September 2021. As a result, we were able to obtain valid content for 347\,010 enterprises. The other home page URLs returned a timeout error or an HTTP code different from 200. Thus, we obtained a valid set of features for the 77\% of the initial sample.
Together with their home page URL, we considered information on firm characteristics, such as geographical localisation (regional - NUTS-2, provincial - NUTS-3 and municipal level), industry (NACE REV.2 digit), age, and size (micro, small, medium or large firm according to the number of employees).
Therefore from the merging of website information with firms' characteristics, we obtain a final sample of 182\.705 observations.

\subsection{Methodology}\label{sec:methodology}

The crawler activity was aimed at extracting different features related to each specific home page to describe: a) how the site is built (i.e., links, images, text, etc.) and b) how it relates to other online contents (e.g., security, link to social media, loading speed, etc.).


We interpreted the economic meaning of each extracted feature elaborating on the existing literature \citep{Mateosetal2001, SandersandGalloway2013, KrolandZdonek2020}. For the present analysis, we considered ten features as follows
\begin{itemize}
\item \textbf{the length of the URL}. Short URLs are easier to remember and are a sign of cleanliness and user-centricity. Moreover, they are more likely to be discovered;
\item \textbf{Social media presence: Facebook, LinkedIn and Instagram}. We consider three popular social media (Facebook, LinkedIn and Instagram), checking for the absence/presence of links to those social media. In fact, social media represent a new valuable tool for conducting digital marketing strategies;
\item \textbf{The quality of internal links}. The presence of unique inner links can be interpreted as a sign of a good level of navigability to increase the probability for users to remain longer on the websites, being correctly addressed towards contents more in line with their preferences;
\item \textbf{The quality of external links}. The higher presence of unique outer links can be interpreted as a strong sign of stakeholder engagement and embeddedness in the digital business ecosystem;
\item \textbf{Quality of the technical frameworks}. The adoption of modern web development standards ensures a better user experience and reflects more technical competencies. This metric is based on Google Lighthouse \footnote{See \texttt{https://github.com/GoogleChrome/lighthouse.}}. 
\item \textbf{Request access time}. A webpage with a low level of request time implies a good speed and a good level of usability, being also this metric at the basis of the adoption of a state-of-the-art technological stack development;
\item \textbf{Website's age}. We estimated the age of each website, checking for its first year of presence in the Wayback Machine archive\footnote{https://archive.org/web/.}. Older websites have been interpreted as part of the digital history and tradition of the firm;
\item \textbf{Website security}. The presence of a high-level security header represents a proxy of awareness towards the risks of cyberattacks. Nowadays, this represents a crucial strategic aspect for firms.
\end{itemize}


In the following, we provide an explorative analysis of the extracted features.  

\begin{figure}[h!]
    \centering
    \begin{tabular}{c}
    \toprule
    \includegraphics[width=0.7\columnwidth]{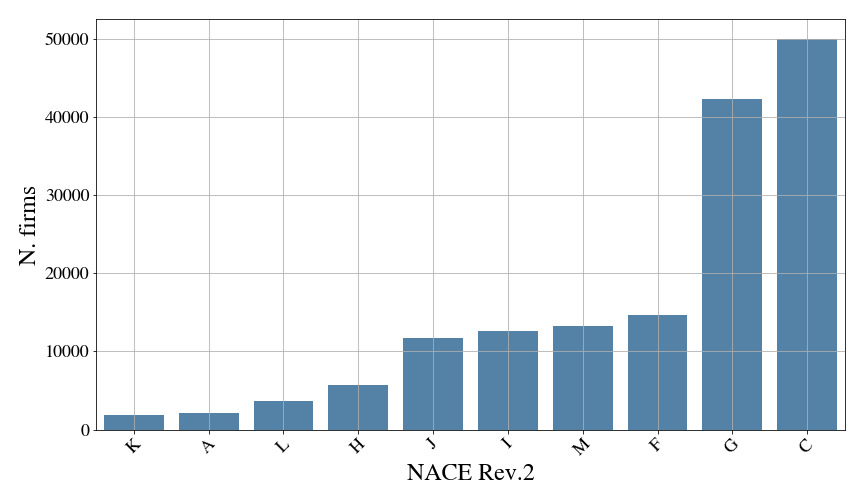} \\
    \includegraphics[width=0.7\columnwidth]{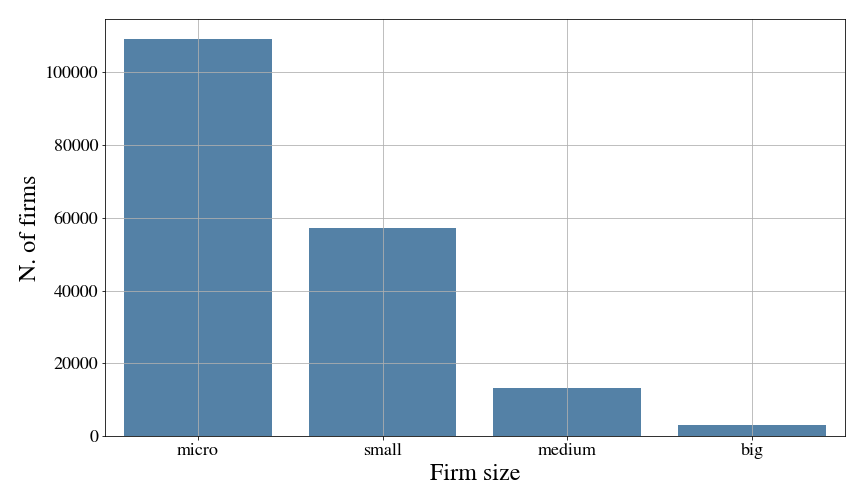} \\
    \includegraphics[width=0.7\columnwidth]{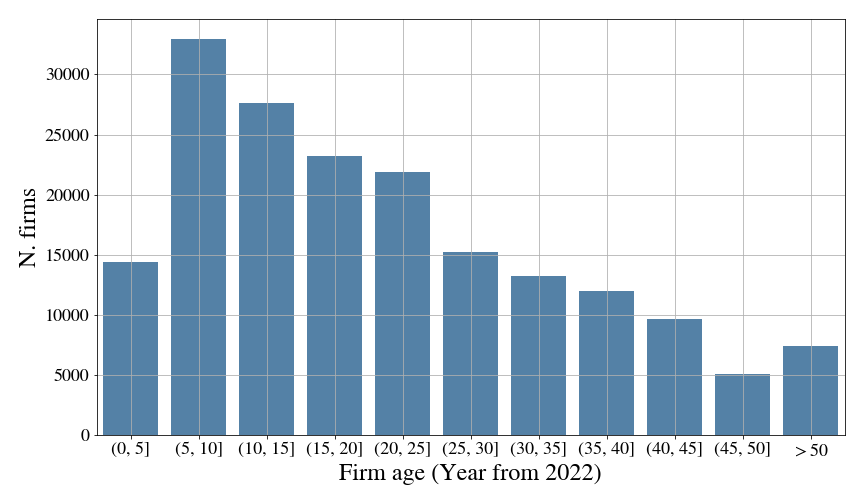} \\
    \bottomrule
    \end{tabular}
    \caption{main features of selected Italian firms with a known physical and digital presence. Top: the number of analysed firms grouped by NACE 2 Digit code, middle: size distribution, Bottom: age distribution.}
    \label{fig:eda_aida}
\end{figure}

Figure \ref{fig:eda_aida} describes how our final dataset is distributed along three firm characteristics, namely industry, size, and age.
On the top, we show the number of enterprises for NACE 2-digit code. In our sample, most of the firms belong to categories C (manufacturing) and G (wholesale and retail trade). 
Notice that not all the industries have been included in this list; in fact, we select only the top 10 more frequent categories.
The second plot shows the size distribution of firms. As expected, considering the known Italian industrial distribution, most of the firms in our dataset are micro and small enterprises, i.e., with less than 50 employees.
At the bottom of figure 3, we report the age of the enterprises in 5-year bins. The results show a higher concentration of relatively young firms that have from 5 to 25 years old.

To investigate the representativeness of the sample, we compare the data we extracted with the general composition of the Italian firms, as reported in ORBIS (Bureau Van Dick) and ISTAT.
Concerning industry, we obtained representativeness that vary from 5\% of NACE sector I (Food and Accommodation) to 28\% of sector J (Information and Communication), with a good level (22\%) also for sector C (Manufacturing) (ISTAT, 2020). 

Concerning size, we were able to capture 75\% of the total number of big firms, with a decreasing coverage for smaller firms: 56\% of medium firms, 31\% of small firms and only 8\% of micro firms (ISTAT, 2020).

In relation to age, we find values in the range 1-6\%,  registering also in this case progressive representativeness going back in time, with the peak (6\%) for the firms born in the period 2002 and 2006 (ORBIS, 2022).

\begin{figure}[ht!]
    \centering
    \footnotesize
    \begin{tabular}{ccc}
    \toprule
    \includegraphics[width=0.25\columnwidth]{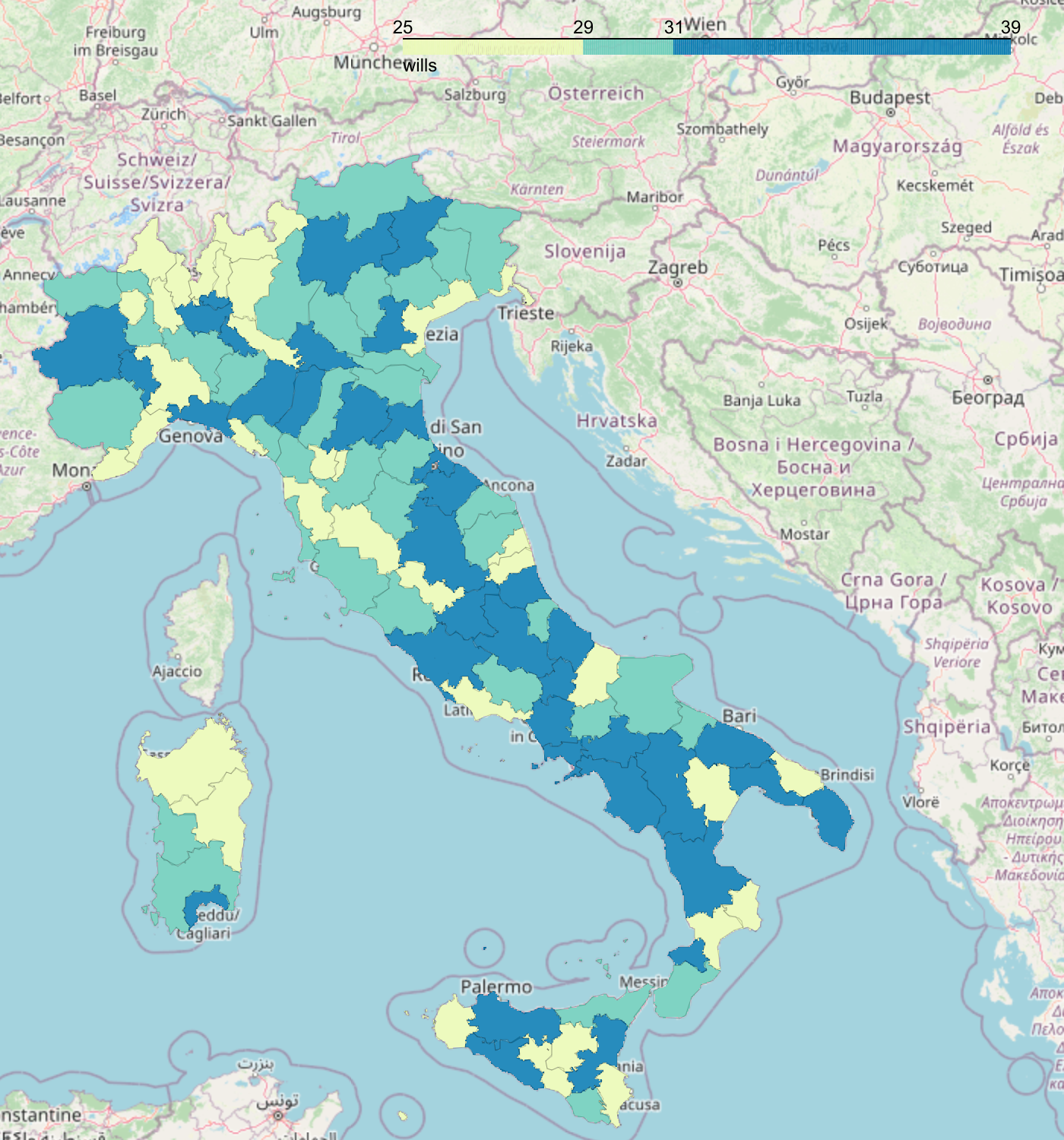} & 
    \includegraphics[width=0.25\columnwidth]{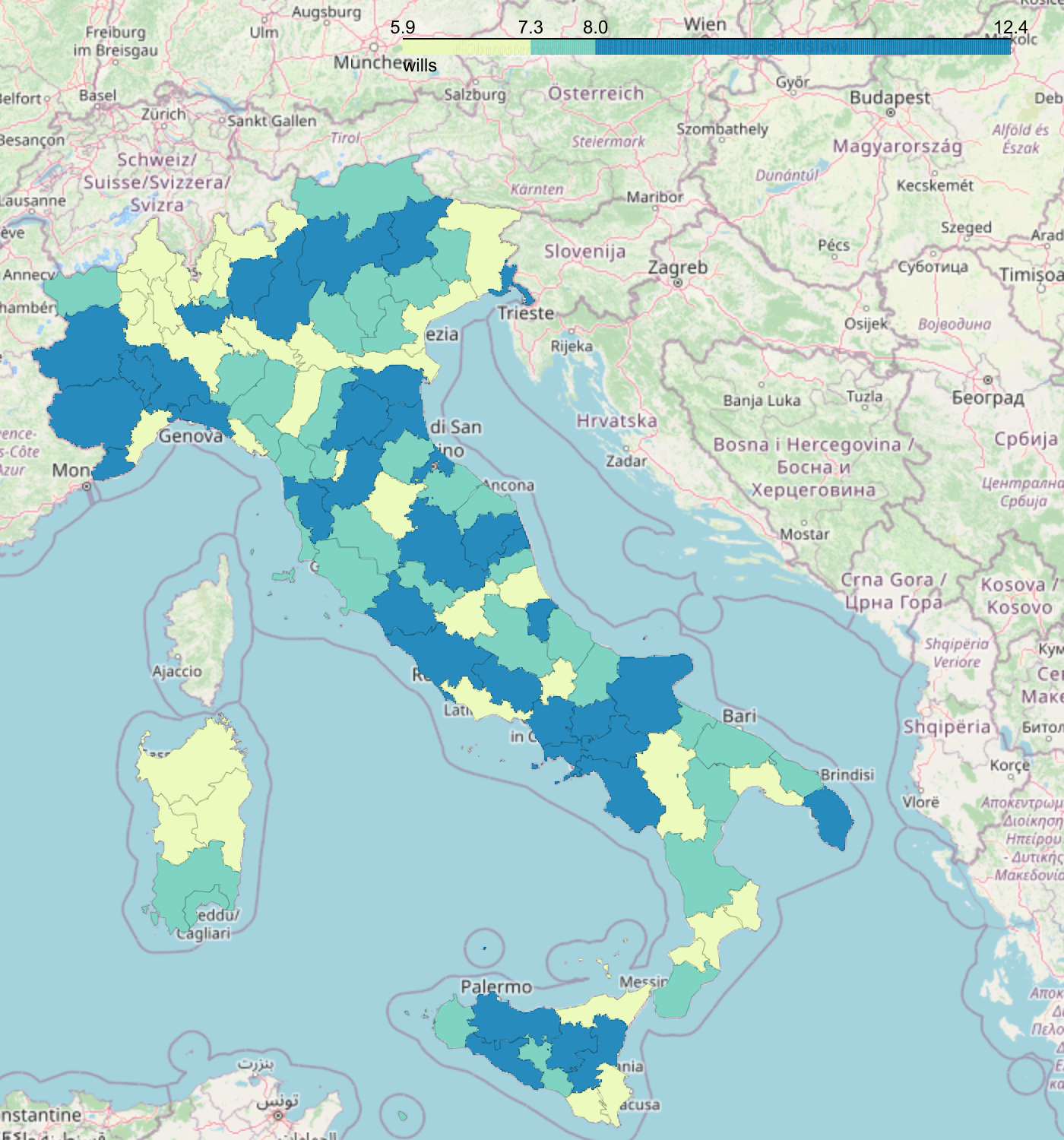} & 
    \includegraphics[width=0.25\columnwidth]{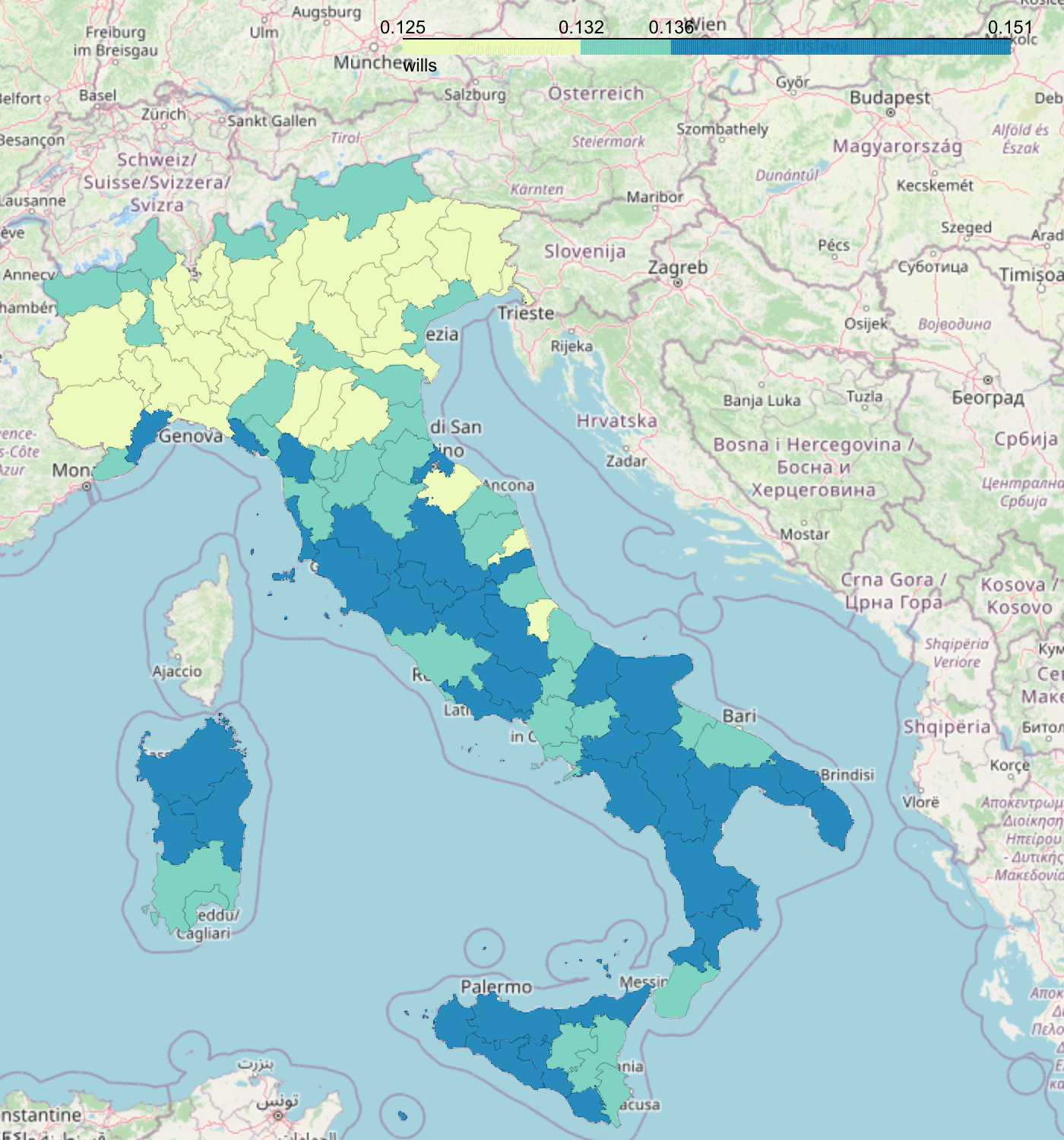}\\
    N.Links IN & N. links out & Lenght URL \\
    \hline
    \includegraphics[width=0.25\columnwidth]{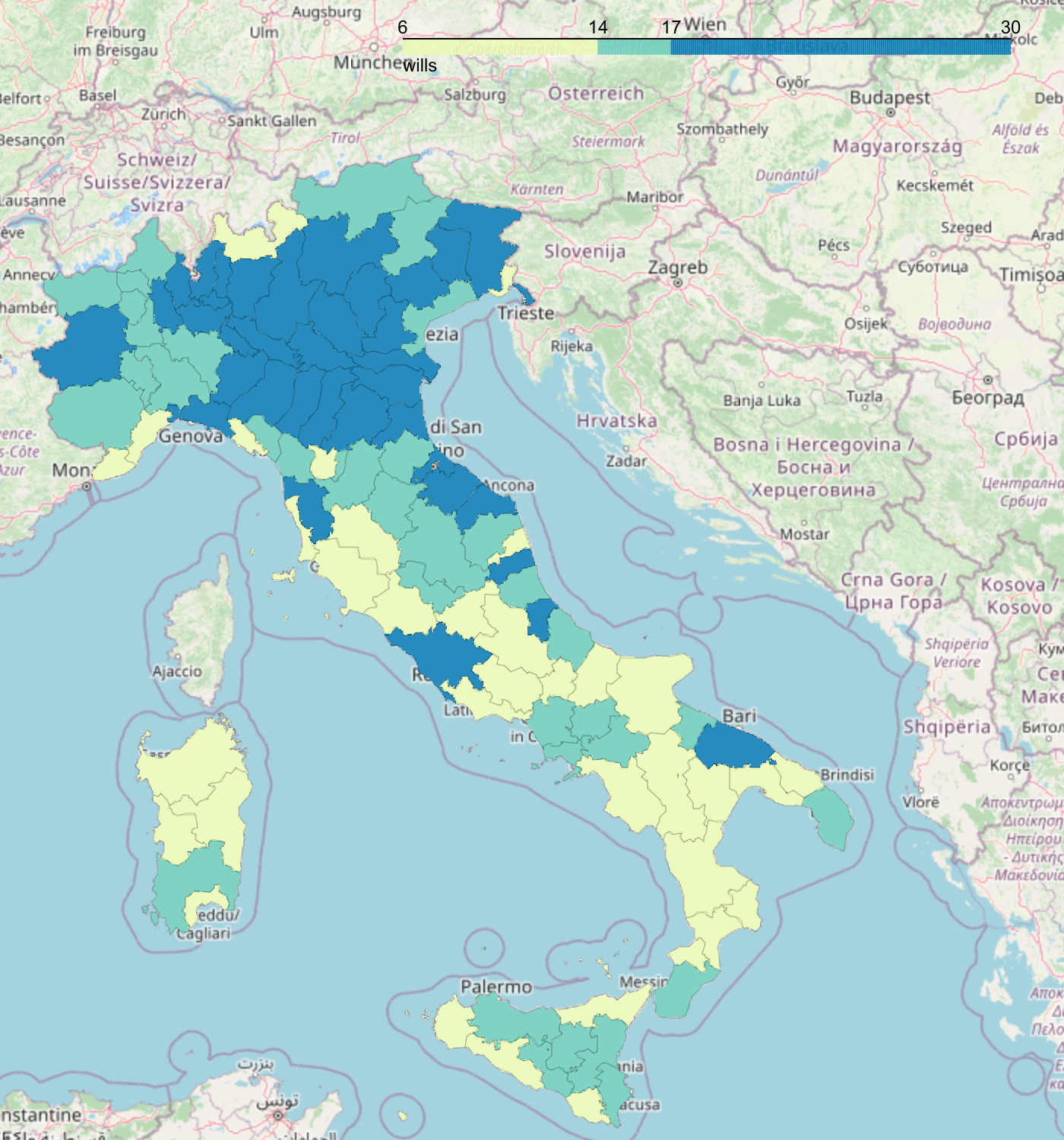} & 
    \includegraphics[width=0.25\columnwidth]{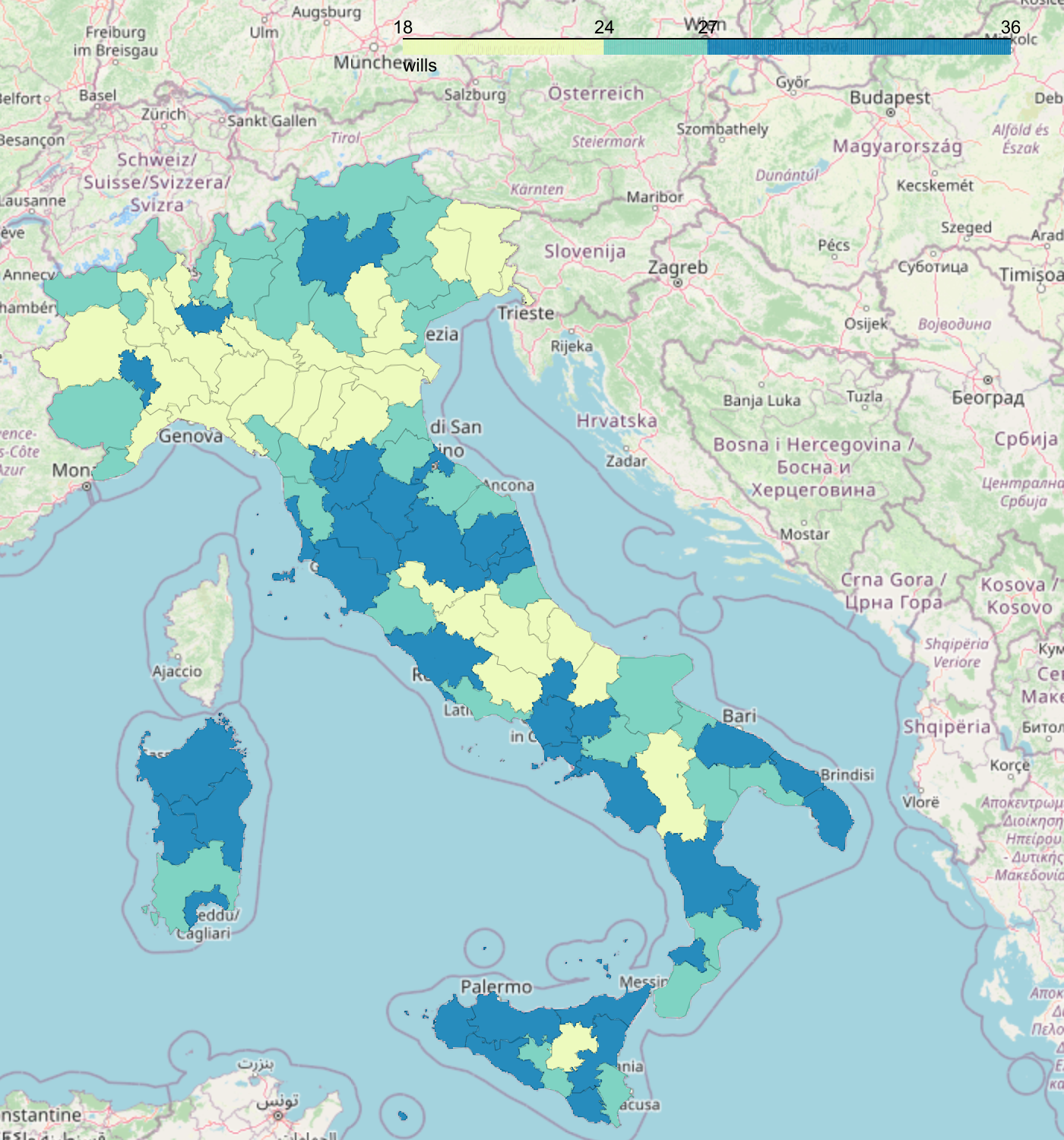} & 
    \includegraphics[width=0.25\columnwidth]{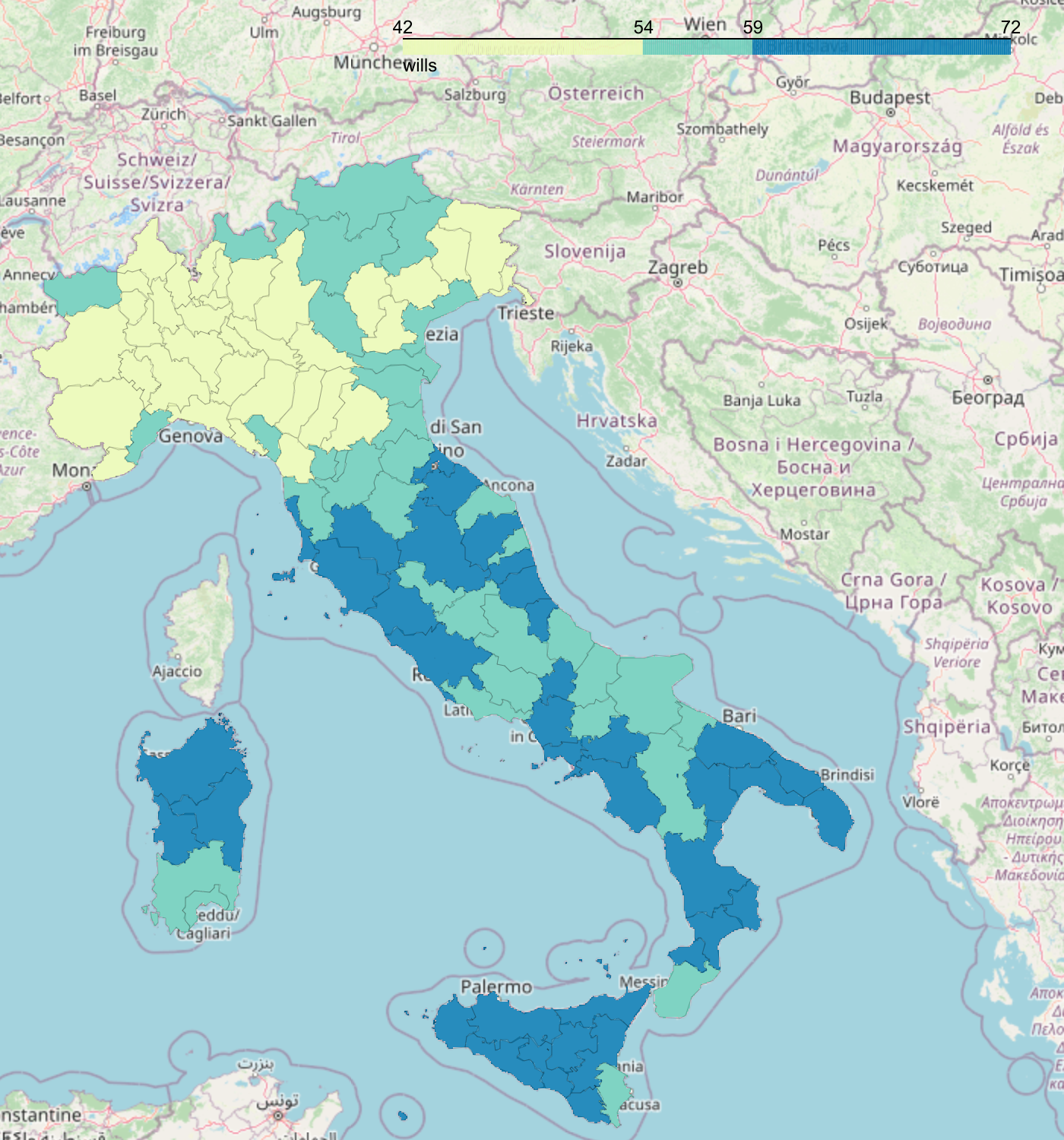}\\
    LinkedIn & Instagram & Facebook \\
    \hline
    \includegraphics[width=0.25\columnwidth]{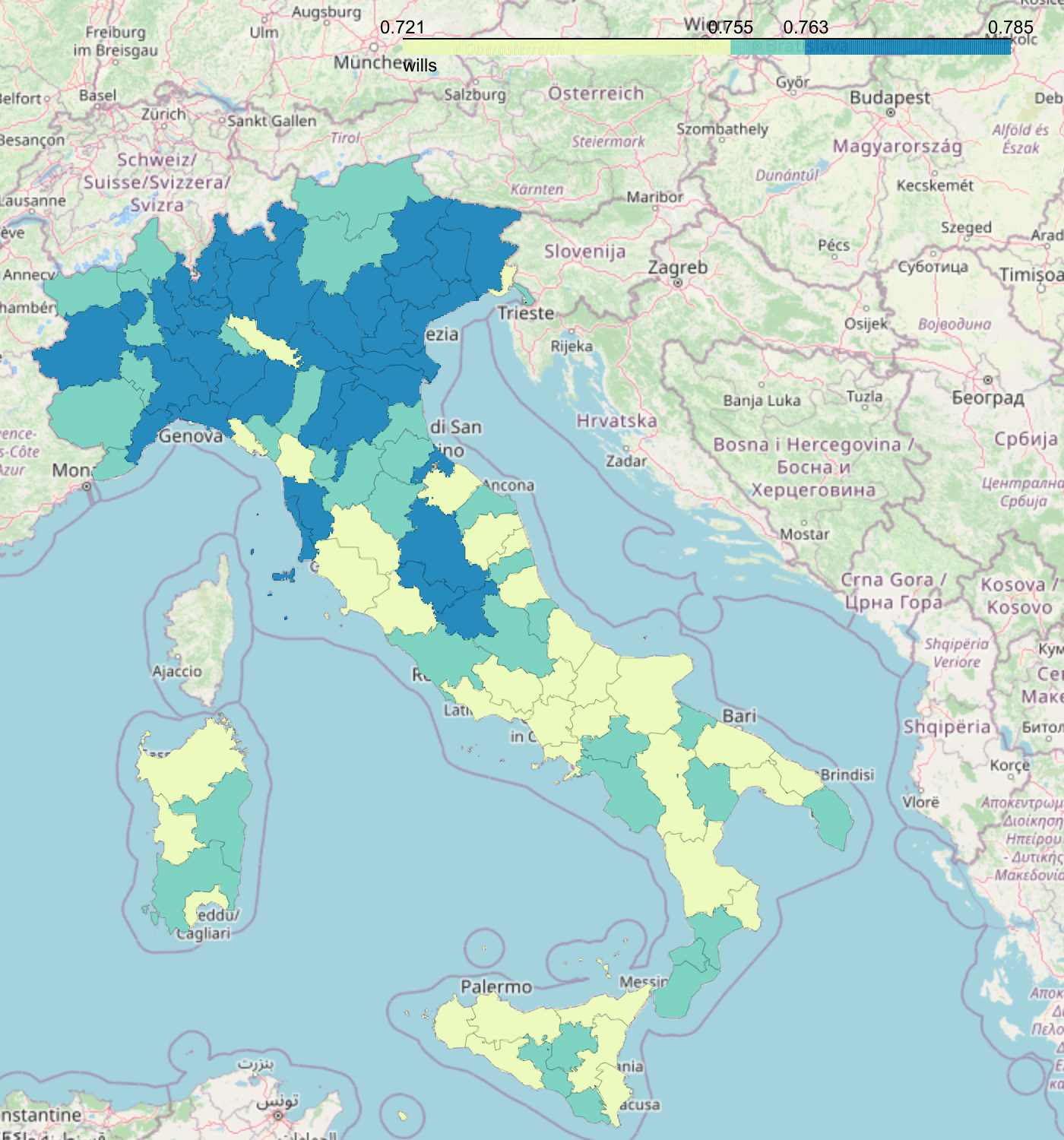} & 
    \includegraphics[width=0.25\columnwidth]{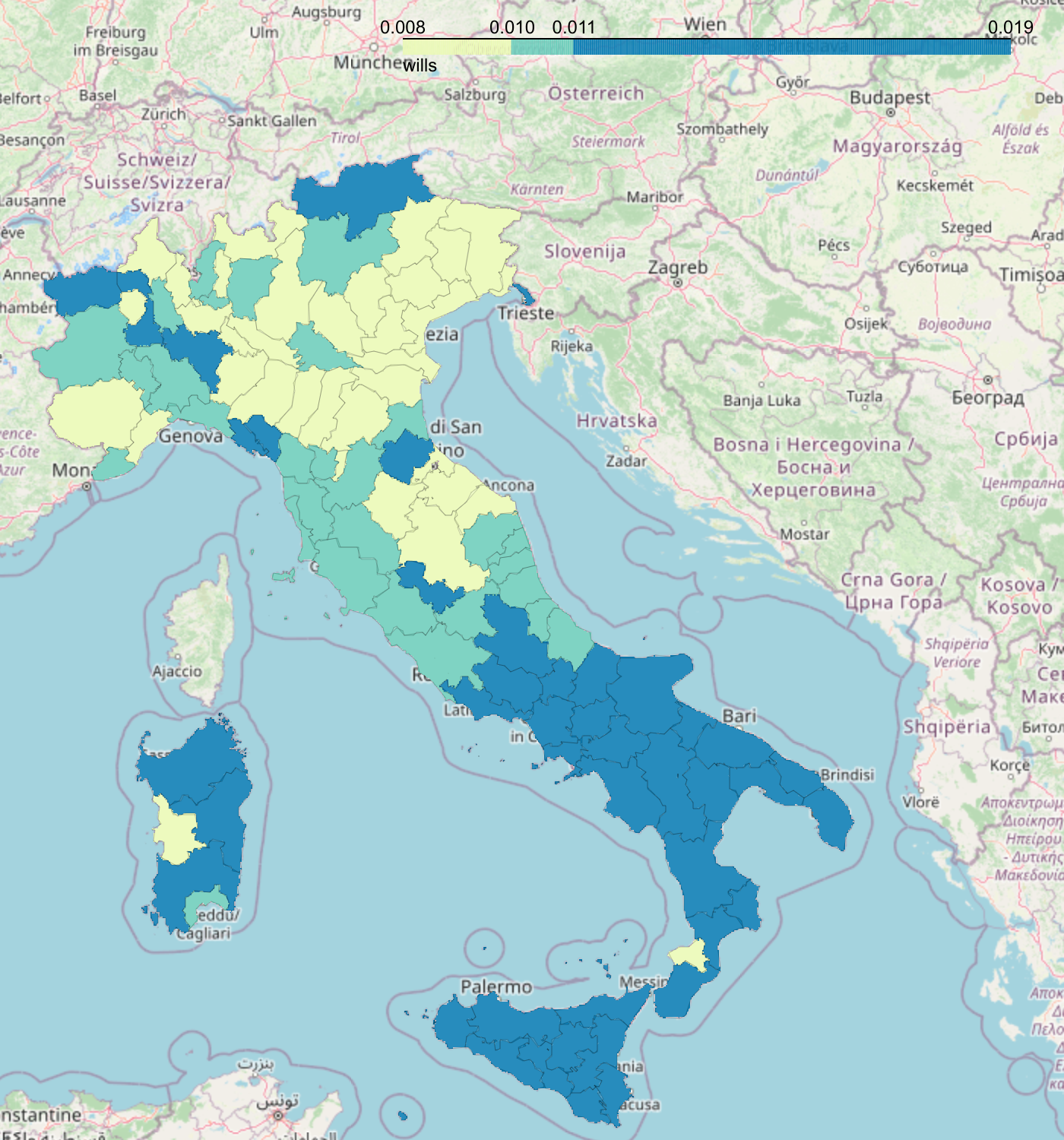} & 
    \includegraphics[width=0.25\columnwidth]{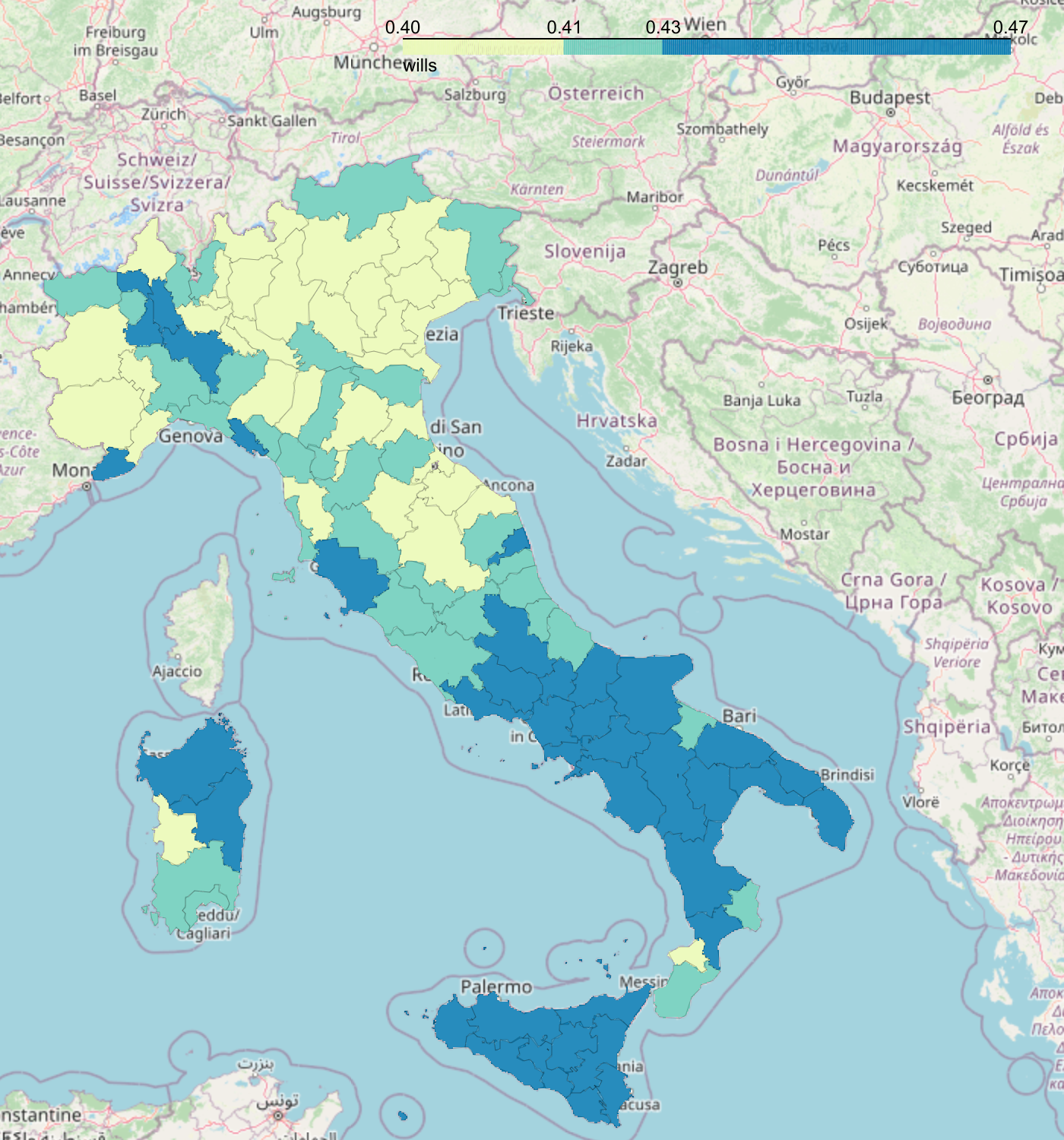}\\
    Best practice & Request access time & Security \\
    \hline
    \multicolumn{3}{c}{
    \begin{tabular}{c}
        \includegraphics[width=0.25\columnwidth]{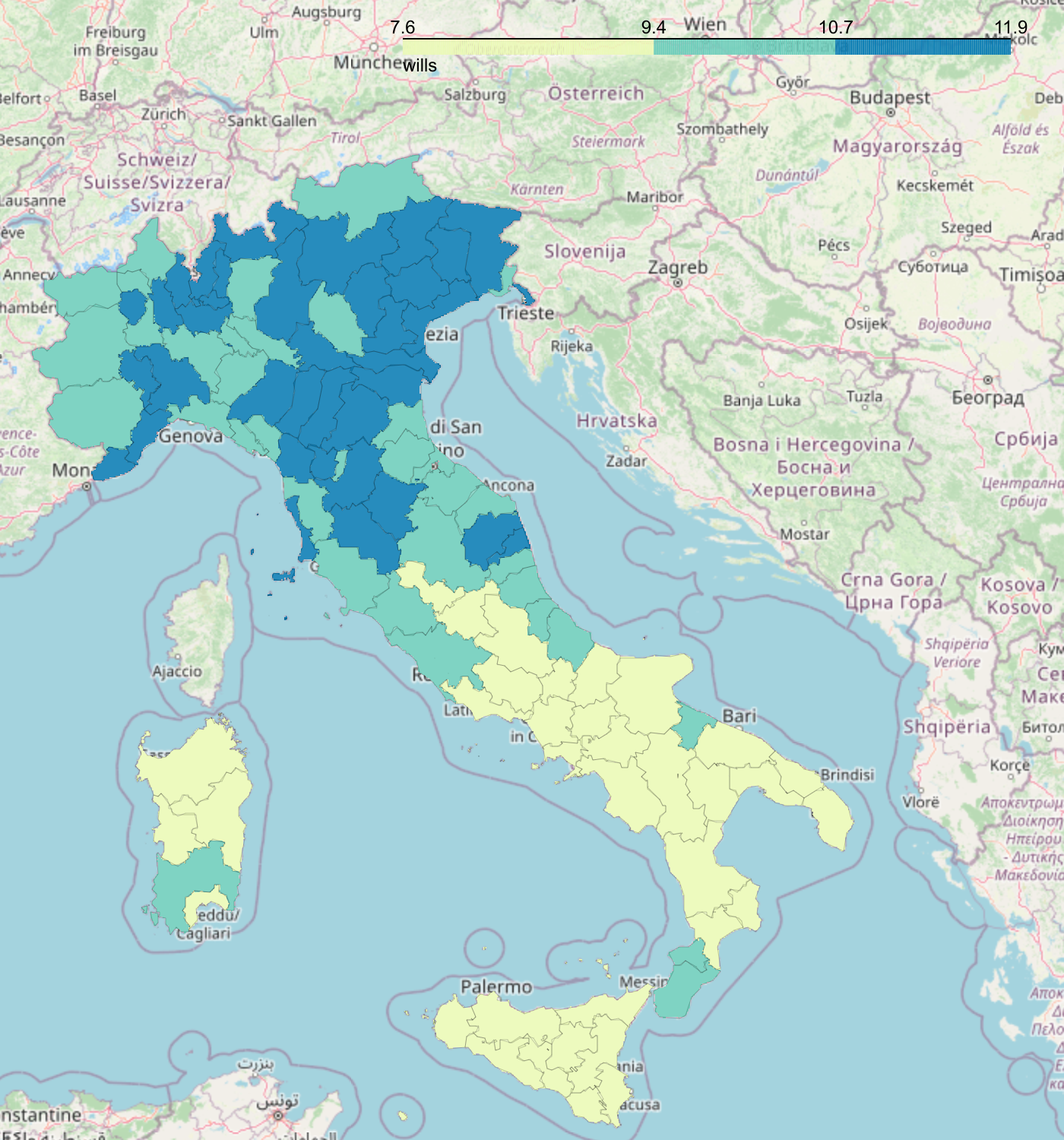} \\
    Website age \\
    \end{tabular}
    }\\
    \bottomrule
    \end{tabular}
    \caption{the spatial distribution of corporate website features in Italy, at the level of NUTS-3 regions (provinces). Features are classified in three quantiles from low (yellow) to high (dark blue) levels.}
    \label{fig:features}
\end{figure}

Figure \ref{fig:features} maps the features extracted by means of the crawling activity at NUTS-3 regions (descriptive statistics are available in table \ref{fig:features}). We report values corresponding to the three quantiles, i.e. 33\%, 66\% and 100\%, where a darker color indicates higher values.
Seven out of the ten extracted features, namely URL length, LinkedIn, quality of the internal links, quality of the technical frameworks, request access time, website age and website security) exhibit clear spatial differences across Italian NUTS-3 regions. 
As an example, The North-South divide in access time might be due to infrastructure problems since a fast broadband connection is more spread in the North than in the South. The spatial distribution of websites might correspond to earlier adoption of digital tools for Northern enterprises w.r.t. firms located in the South, thus indicating a more rooted digital mindset.
More scattered territorial patterns are reported for external links and social media. Social media are of special interest for their different penetration degree revealed, suggesting a specific interpretation of their usage.
The presence of Facebook can be interpreted as the diffusion of a more generic marketing culture, with a low level of specialisation. Instagram can be seen as the diffusion of specific marketing culture adopted by specific B2C industries that exploit images and videos (such as tourism, food, and culture). LinkedIn can be interpreted as the diffusion of Human Resources culture and a proxy of more developed job markets, where relying on professional networks can support more concrete business activities.


\begin{table}[]
    \centering
\begin{tabular}{lrrrr}
\toprule
{} &   min &       max &   mean &    std \\
\midrule
unique\_links\_in     & 0.000 &  4885.000 & 27.473 & 48.921 \\
unique\_links\_out    & 0.000 & 15814.000 &  6.942 & 42.228 \\
best-practices      & 0.380 &     1.000 &  0.837 &  0.103 \\
length\_url          & 4.000 &   105.000 & 19.214 &  5.436 \\
Facebook            & 0.000 &     1.000 &  0.461 &  0.498 \\
Instagram           & 0.000 &     1.000 &  0.219 &  0.414 \\
LinkedIn            & 0.000 &     1.000 &  0.159 &  0.366 \\
years\_old           & 0.000 &    25.000 & 10.308 &  7.455 \\
request\_time        & 0.127 &   502.262 &  5.265 &  9.418 \\
security\_header\_int & 0.000 &    15.000 &  6.268 &  1.941 \\
\bottomrule
\end{tabular}
    \caption{Descriptive statistics of the features of the corporate websites of interest}
    \label{tab:feature_description}
\end{table}

Interestingly, table \ref{tab:correlation_matrix} shows a low level of correlation between variables, with the only exception of correlation between social media (ca .5). This result supports our approach of collecting different indicators to capture a multifaceted topic such as the corporate digital divide.

\begin{sidewaystable}
\centering
\begin{adjustbox}{max width=1.\textwidth,center}
\begin{tabular}{lrrrrrrrrrr}
\toprule
{} &  unique\_links\_in &  unique\_links\_out &  best-practices &  Instagram &  LinkedIn &  Facebook &  years\_old &  security\_header\_int &  request\_time &  length\_url \\
\midrule
unique\_links\_in     &            1.000 &             0.018 &          -0.042 &      0.151 &     0.104 &     0.179 &      0.078 &               -0.053 &        -0.009 &      -0.023 \\
unique\_links\_out    &            0.018 &             1.000 &          -0.005 &      0.073 &     0.062 &     0.087 &      0.022 &               -0.013 &         0.001 &      -0.003 \\
best-practices      &           -0.042 &            -0.005 &           1.000 &      0.038 &     0.055 &     0.008 &     -0.038 &               -0.055 &        -0.011 &      -0.026 \\
Instagram           &            0.151 &             0.073 &           0.038 &      1.000 &     0.178 &     0.524 &      0.007 &               -0.032 &         0.006 &      -0.004 \\
LinkedIn            &            0.104 &             0.062 &           0.055 &      0.178 &     1.000 &     0.293 &      0.080 &               -0.074 &        -0.040 &      -0.096 \\
Facebook            &            0.179 &             0.087 &           0.008 &      0.524 &     0.293 &     1.000 &      0.002 &               -0.033 &         0.001 &       0.023 \\
years\_old           &            0.078 &             0.022 &          -0.038 &      0.007 &     0.080 &     0.002 &      1.000 &               -0.056 &        -0.060 &      -0.167 \\
security\_header\_int &           -0.053 &            -0.013 &          -0.055 &     -0.032 &    -0.074 &    -0.033 &     -0.056 &                1.000 &         0.769 &       0.072 \\
request\_time        &           -0.009 &             0.001 &          -0.011 &      0.006 &    -0.040 &     0.001 &     -0.060 &                0.769 &         1.000 &       0.075 \\
length\_url          &           -0.023 &            -0.003 &          -0.026 &     -0.004 &    -0.096 &     0.023 &     -0.167 &                0.072 &         0.075 &       1.000 \\
\bottomrule
\end{tabular}
\end{adjustbox}
    \caption{Correlation matrix between the ten corporate website features}
    \label{tab:correlation_matrix}

\end{sidewaystable}




More in general, our explorative data analysis allows us to validate some characteristics of the enterprises included in our dataset built through the crawler. As a result, we rely on 182\,705 firms out of the initial sample (composed of 347\,010 firms) because a consistent number of firms showed missing values for employment, a useful indicator to categorize the size of each firm in different groups. To analyse the relationship between firms' characteristics and corporate websites we run a set of simple OLS regressions with the following specification:
 
\begin{equation*}
    y = \alpha + \beta_1x_f +  \beta_2x_t + \epsilon
\end{equation*}
where $y$ represents one of the ten Digital Dimensions (i.e., one of the relevant features of the corporate websites), and $x_f$ is the vector that represents the relevant characteristics of the firm: age, industry and size. In each regression, we include a set of controls for the area in which the firm is located, represented by $x_t$. This is because, beyond the specific patterns detectable by the traditional characteristics of the firm, the notion of corporate Digital Divide can be influenced (and captured) also by the place where the firm is located (e.g. a common available IT infrastructure or the diffusion of digital means to run a business activity varies between urban and rural areas and from the North to the South). In this regard, to avoid possible mispecifications of the model, (since some features of corporate websites, such as the request access time might depend on the wide band coverage of the area where the firm is located), we control for the possible impact of the wide band. 
The wide band data are obtained as open data from the Italian Authority for Communications Guarantees\footnote{https://maps.agcom.it/}. The data represent, 
for each Italian census area, the ratio of households reached by broadband with a speed greater than 30mbs over the total number of households in that area.
\section{Results}\label{sec:regression}

\subsection{Regression results}

In this section, we discuss the results of the regression analysis for the ten digital-related features of corporate websites.
Table \ref{tab:regression_features} summarizes the results for the analysis of the Digital Divide across different dimensions of Italian firms: size, sector, age and location.

Micro firms show a negative and significant effect for unique links internal to the company, with more attention to internal navigability shown by medium and big firms. External links are not significant in micro firms, while we find a positive and significant effect for medium and large firms as a sign of more attention devoted to external stakeholders in terms of connection and reachability.
Best practice has a positive and significant effect only on medium-sized firms, with no effect on big firms. This counterintuitive result could be interpreted as a sign of lower marginal utility in having libraries at the state of the art for big firms.

Social media variables have all a significant effect across dimensions, with a negative sign for micro and a progressive value of the magnitude effect, with the highest value for big firms.

Age, as a proxy of digital experience, tends to be more relevant in big firms, which, on average, understood the importance of creating a website many years before.

Security, speed (request time), and the length of the URL should be interpreted in the inverted direction as lower request times and shorter URL improve the accessibility of corporate websites.


We analyze the impact of the corporate Digital Divide also across NUTS-2 digit sectors.
We take as a sector of reference in terms of quality, Information and Communication (``J''), for its direct involvement in the coding procedure or the marked attention to the quality of websites, as a fundamental tool to communicate with customers. 
A positive and significant effect is found in the wholesale and retail trade (``G'') for the unique internal and external links, confirming the importance of internal navigability and external connections for B2C activities.
Concerning best practice, a positive and significant effect is detected in the sectors ``M'' (Professional, Scientific and Technical activities and ``J'' (Information and Communication), supporting the idea that, on average, firms in those sectors are better informed about the importance of high-quality technical libraries. Social Media seems to have different behaviors, according to their business functions, confirming the exploratory mapping reported in section \ref{sec:data}. Accordingly, Instagram and Facebook have seems to play an important role in Accommodation and Food Service Activities (sector ``I''), as free channels to engage with tourism. LinkedIn, more oriented to attracting highly skilled profiles, seems to have a bigger impact on sectors with highly specialized knowledge.

Security Header and request time should be interpreted in the opposite direction, so negative and significant effects are signs of firms being more aware of the importance of adopting cyber security practices and frameworks able to ensure an adequate level of speed.



Age accounts for the years since the date of foundation. A negative and significant effect for unique links, best practices and social media is a sign that older firms have a worse performance than more digital-born companies.

The geography of the corporate Digital Divide also presents sharp evidence of the fact that urban contexts offer the higher marginal utility of digital means usage for firms operating in dynamic and dense economic systems, where firms can exploit the network effect \citep{Formanetal2005}.
The geographical location is also another important feature that reveals how the Digital Divide is the reverse coin of the economic development of territories. North and South dummies show opposite behavior. This supports the idea that more industrialized regions, prevalently located in the north, are favorable environments for the adoption of digital strategies and the relative adoption of skills and investment more than in the south.

In all the regressions we control for the effect of the wide band, supposing that some digital web proxies could be shadowed by a high-quality web infrastructure.
Disentangling the infrastructural effect is not a trivial task. Accordingly, websites can rely on external hosts or local servers.
In the first case, the reactivity of the different webpages can be related to the quality of the website.
In the last case, we can observe a good level of wide band coverage with low-quality websites, but also the opposite might occur. All in all, we find only four significant values for the impact of wide band out of ten regressions. This result is very important since it confirms that with our approach to collect multiple features of corporate websites we are able to detect the digital footprint of firms beyond the quality of shared web infrastructure. 


\begin{sidewaystable}
\centering
\scriptsize
\begin{adjustbox}{max width=1.\textwidth,center}
\begin{tabular}{lcccccccccc}
\toprule
{} & unique\_links\_in &               unique\_links\_out &                 best-practices &                      Instagram &                       LinkedIn &                       Facebook &                      years\_old &            security\_header\_int &                   request\_time &                     length\_url \\
\midrule
Constant &   \makecell{ 0.005*** \\(0.000)} &   \makecell{ 0.000*** \\(0.000)} &   \makecell{ 0.771*** \\(0.002)} &  \makecell{ -0.965*** \\(0.027)} &  \makecell{ -1.810*** \\(0.032)} &   \makecell{ 0.376*** \\(0.022)} &   \makecell{ 0.406*** \\(0.003)} &   \makecell{ 0.415*** \\(0.001)} &   \makecell{ 0.010*** \\(0.000)} &   \makecell{ 0.143*** \\(0.000)} \\
Micro firm &  \makecell{ -0.001*** \\(0.000)} &  \makecell{ -0.000*** \\(0.000)} &  \makecell{ -0.005*** \\(0.001)} &  \makecell{ -0.236*** \\(0.013)} &  \makecell{ -0.576*** \\(0.016)} &  \makecell{ -0.136*** \\(0.011)} &  \makecell{ -0.093*** \\(0.002)} &   \makecell{ 0.012*** \\(0.001)} &   \makecell{ 0.001*** \\(0.000)} &   \makecell{ 0.002*** \\(0.000)} \\
Mid-sized firm &   \makecell{ 0.002*** \\(0.000)} &   \makecell{ 0.000*** \\(0.000)} &   \makecell{ 0.006*** \\(0.001)} &   \makecell{ 0.322*** \\(0.023)} &   \makecell{ 0.654*** \\(0.023)} &   \makecell{ 0.196*** \\(0.020)} &   \makecell{ 0.074*** \\(0.003)} &  \makecell{ -0.013*** \\(0.001)} &  \makecell{ -0.001*** \\(0.000)} &  \makecell{ -0.002*** \\(0.000)} \\
Large firm &   \makecell{ 0.005*** \\(0.000)} &   \makecell{ 0.000*** \\(0.000)} &      \makecell{ 0.003 \\(0.003)} &   \makecell{ 0.505*** \\(0.042)} &   \makecell{ 1.113*** \\(0.040)} &   \makecell{ 0.339*** \\(0.038)} &   \makecell{ 0.109*** \\(0.005)} &  \makecell{ -0.029*** \\(0.002)} &     \makecell{ -0.001 \\(0.000)} &   \makecell{ -0.002** \\(0.001)} \\
Agriculture, forestry and fishing (A) &   \makecell{ -0.001** \\(0.000)} &     \makecell{ -0.000 \\(0.000)} &      \makecell{ 0.001 \\(0.003)} &   \makecell{ 0.694*** \\(0.049)} &  \makecell{ -0.575*** \\(0.090)} &   \makecell{ 0.260*** \\(0.047)} &   \makecell{ 0.031*** \\(0.007)} &    \makecell{ -0.007* \\(0.003)} &     \makecell{ -0.000 \\(0.000)} &   \makecell{ -0.003** \\(0.001)} \\
Manufacturing (C) &  \makecell{ -0.000*** \\(0.000)} &   \makecell{ -0.000** \\(0.000)} &     \makecell{ -0.001 \\(0.001)} &  \makecell{ -0.134*** \\(0.020)} &   \makecell{ 0.272*** \\(0.023)} &  \makecell{ -0.494*** \\(0.016)} &   \makecell{ 0.099*** \\(0.002)} &    \makecell{ -0.002* \\(0.001)} &   \makecell{ -0.000** \\(0.000)} &  \makecell{ -0.015*** \\(0.000)} \\
Construction (F) &  \makecell{ -0.001*** \\(0.000)} &  \makecell{ -0.000*** \\(0.000)} &     \makecell{ -0.002 \\(0.002)} &  \makecell{ -0.804*** \\(0.031)} &  \makecell{ -0.194*** \\(0.033)} &  \makecell{ -0.653*** \\(0.022)} &  \makecell{ -0.012*** \\(0.003)} &   \makecell{ 0.010*** \\(0.001)} &   \makecell{ 0.001*** \\(0.000)} &  \makecell{ -0.004*** \\(0.000)} \\
Wholesale and retail trade (G) &   \makecell{ 0.003*** \\(0.000)} &   \makecell{ 0.000*** \\(0.000)} &  \makecell{ -0.005*** \\(0.001)} &   \makecell{ 0.300*** \\(0.019)} &     \makecell{ -0.002 \\(0.024)} &     \makecell{ 0.033* \\(0.016)} &   \makecell{ 0.061*** \\(0.002)} &      \makecell{ 0.001 \\(0.001)} &   \makecell{ 0.001*** \\(0.000)} &  \makecell{ -0.010*** \\(0.000)} \\
Transportation (H) &  \makecell{ -0.002*** \\(0.000)} &     \makecell{ -0.000 \\(0.000)} &  \makecell{ -0.009*** \\(0.002)} &  \makecell{ -0.763*** \\(0.044)} &  \makecell{ -0.287*** \\(0.046)} &  \makecell{ -0.735*** \\(0.031)} &   \makecell{ 0.022*** \\(0.004)} &   \makecell{ 0.007*** \\(0.002)} &      \makecell{ 0.000 \\(0.000)} &  \makecell{ -0.003*** \\(0.001)} \\
Accomodation and food service (I) &  \makecell{ -0.001*** \\(0.000)} &      \makecell{ 0.000 \\(0.000)} &  \makecell{ -0.015*** \\(0.002)} &   \makecell{ 0.904*** \\(0.024)} &  \makecell{ -1.411*** \\(0.052)} &   \makecell{ 0.741*** \\(0.023)} &   \makecell{ 0.045*** \\(0.003)} &     \makecell{ -0.002 \\(0.001)} &  \makecell{ -0.001*** \\(0.000)} &   \makecell{ 0.009*** \\(0.001)} \\
Information and communication (J) &   \makecell{ 0.001*** \\(0.000)} &   \makecell{ 0.000*** \\(0.000)} &    \makecell{ 0.005** \\(0.002)} &  \makecell{ -0.262*** \\(0.029)} &   \makecell{ 1.275*** \\(0.028)} &  \makecell{ -0.133*** \\(0.023)} &   \makecell{ 0.070*** \\(0.003)} &  \makecell{ -0.019*** \\(0.001)} &  \makecell{ -0.002*** \\(0.000)} &  \makecell{ -0.027*** \\(0.001)} \\
Financial and insurance activities (K) &    \makecell{ -0.001* \\(0.000)} &     \makecell{ -0.000 \\(0.000)} &      \makecell{ 0.003 \\(0.004)} &  \makecell{ -0.982*** \\(0.080)} &   \makecell{ 0.723*** \\(0.059)} &  \makecell{ -0.709*** \\(0.051)} &     \makecell{ 0.017* \\(0.007)} &  \makecell{ -0.017*** \\(0.003)} &  \makecell{ -0.003*** \\(0.000)} &  \makecell{ -0.009*** \\(0.001)} \\
Real estate activities (L) &      \makecell{ 0.001 \\(0.000)} &     \makecell{ 0.000* \\(0.000)} &  \makecell{ -0.011*** \\(0.003)} &   \makecell{ 0.168*** \\(0.043)} &   \makecell{ 0.263*** \\(0.052)} &      \makecell{ 0.036 \\(0.036)} &   \makecell{ 0.084*** \\(0.005)} &     \makecell{ -0.002 \\(0.002)} &     \makecell{ -0.001 \\(0.000)} &      \makecell{ 0.001 \\(0.001)} \\
Professional, scientific \& technical activities (M) &     \makecell{ -0.000 \\(0.000)} &      \makecell{ 0.000 \\(0.000)} &   \makecell{ 0.008*** \\(0.002)} &  \makecell{ -0.212*** \\(0.027)} &   \makecell{ 1.137*** \\(0.027)} &  \makecell{ -0.385*** \\(0.022)} &   \makecell{ 0.039*** \\(0.003)} &  \makecell{ -0.014*** \\(0.001)} &  \makecell{ -0.002*** \\(0.000)} &  \makecell{ -0.018*** \\(0.000)} \\
Urban area &   \makecell{ 0.000*** \\(0.000)} &    \makecell{ 0.000** \\(0.000)} &      \makecell{ 0.001 \\(0.001)} &   \makecell{ 0.130*** \\(0.013)} &   \makecell{ 0.170*** \\(0.015)} &   \makecell{ 0.052*** \\(0.011)} &      \makecell{ 0.003 \\(0.002)} &     \makecell{ -0.002 \\(0.001)} &     \makecell{ -0.000 \\(0.000)} &  \makecell{ -0.002*** \\(0.000)} \\
North &      \makecell{ 0.000 \\(0.000)} &     \makecell{ -0.000 \\(0.000)} &   \makecell{ 0.011*** \\(0.001)} &  \makecell{ -0.102*** \\(0.015)} &   \makecell{ 0.288*** \\(0.018)} &  \makecell{ -0.201*** \\(0.012)} &   \makecell{ 0.019*** \\(0.002)} &  \makecell{ -0.008*** \\(0.001)} &  \makecell{ -0.001*** \\(0.000)} &  \makecell{ -0.005*** \\(0.000)} \\
South &   \makecell{ 0.000*** \\(0.000)} &     \makecell{ -0.000 \\(0.000)} &  \makecell{ -0.008*** \\(0.001)} &   \makecell{ 0.067*** \\(0.018)} &  \makecell{ -0.109*** \\(0.023)} &   \makecell{ 0.162*** \\(0.015)} &  \makecell{ -0.045*** \\(0.002)} &   \makecell{ 0.014*** \\(0.001)} &   \makecell{ 0.002*** \\(0.000)} &    \makecell{ 0.001** \\(0.000)} \\
Firm age &  \makecell{ -0.000*** \\(0.000)} &      \makecell{ 0.000 \\(0.000)} &  \makecell{ -0.000*** \\(0.000)} &  \makecell{ -0.013*** \\(0.000)} &  \makecell{ -0.009*** \\(0.000)} &  \makecell{ -0.010*** \\(0.000)} &                           &    \makecell{ 0.000** \\(0.000)} &     \makecell{ -0.000 \\(0.000)} &   \makecell{ 0.000*** \\(0.000)} \\
Wide band &    \makecell{ 0.000** \\(0.000)} &      \makecell{ 0.000 \\(0.000)} &      \makecell{ 0.001 \\(0.001)} &     \makecell{ 0.043* \\(0.017)} &   \makecell{ 0.131*** \\(0.020)} &      \makecell{ 0.015 \\(0.014)} &    \makecell{ -0.004* \\(0.002)} &      \makecell{ 0.002 \\(0.001)} &      \makecell{ 0.000 \\(0.000)} &     \makecell{ -0.000 \\(0.000)} \\
\midrule
\midrule
R-squared: &                          0.028 &                          0.002 &                          0.005 &              0.037               &        	0.078                     &             0.035                &                          0.071 &                          0.013 &                          0.009 &                          0.041 \\
Adj. R-squared: &                          0.027 &                          0.002 &                          0.005 &                             &                             &                             &                          0.071 &                          0.013 &                          0.009 &                          0.041 \\
\midrule
N. of Observations & 182,705 & 182,705 & 182,705 & 182,705 & 182,705 & 182,705 & 182,705 &182,705 &182,705 &182,705 \\

\bottomrule

\end{tabular}

\end{adjustbox}
\caption{\label{tab:regression_features} regression results considering as dependent variables the website feature extracted. Notice that for the features Instagram, Linkedin and Facebook, we use the Logit model and consequently, we report the relative Pseudo R-squared.}
\end{sidewaystable}

\subsection{Corporate Digital Assessment Index}

Our findings show a multifaceted composite picture, revealing how Digital Divide is present across different firms' dimensions.

A major problem in measuring the corporate Digital Divide with such a data-driven approach is the absence of a theoretical framework to classify the typology of proxies into well-defined digital capabilities.

A very specific research stream narrows on this, analysing the quality of websites as a proxy of organisations' behavior \citep{Mateosetal2001, Goketal2015, AxenbeckBreithaupt2019, KinneandAxenback2020}.
Analysing the quality of websites is a complex task for the presence of several unstructured pieces of information that need to be framed into logical theoretical schemes to understand their potential economic value.
Moreover, it is not trivial to find synthetic measures able to capture the multifaceted characteristics of websites, considering also how a high degree of customisation makes it difficult to establish "objective" accountable criteria \citep{KrolandZdonek2020}.


Within this literature, \cite{Mateosetal2001} provide one of the first single index, called Web Assessment Index (WAI).\footnote{The authors built their index on universities' websites, considering four dimensions of analysis: site content, speed, accessibility and navigability. While the first two are quite straightforward, the last two need to be explicitly framed. As underlined also by \cite{Vaughan2004}, accessibility has been proxied respectively by search engine indexes and popularity (the number of external links can imply more traffic). Navigability has been mainly measured as the number of steps (clicks) to access relevant information for the user. Other measures have been introduced as usability, which can be seen as an extension of navigability, more oriented to the convenience of users to navigate the website for several reasons, such as easiness, responsivity, and aesthetic reasons, which have become crucial especially in the last years to retain and acquire users \citep{Mateosetal2001, DickingerandStangl2013, KrolandZdonek2020}.}

Elaborating on this literature from the point of view of the digital strategy of the firm, we interpreted and classified the extracted features into four different aspects related to the digital strategy of the firm: 

\begin{enumerate}
    \item stakeholder engagement
    \item technical capabilities
    \item internal organisation
    \item digital culture
\end{enumerate}

Stakeholder engagement includes the capabilities of the firm to establish links and connections with external stakeholders, such as customers and suppliers. Technical capabilities account for the competencies of the firm to build a state-of-the-art digital framework in terms of Libraries, security and speed. Internal organisation represents the ability of the firm to coherently and purposefully orchestrate internal information flows. Digital Culture stands for the  historical trajectory in which the firm is inserted (awareness of the importance of web as a strategic resource). 

Aggregating these four indicators, we propose a synthetic index to capture the Corporate digital divide with a single dimension.

To build our index, we normalize the variables with a $MinMax$ technique, inverting the scale of the values where necessary (the length of the URL, Facebook, request time, and security header) for the purpose of interpretability. Then, we calculated the Corporate Digital Assessment Index (CoDAI) as the weighted sum of each element, as follows:

\begin{equation}
\begin{adjustbox}{max width=1.\textwidth,center}

\label{eq:wai}
CoDAI = (Stakeholder  Engagement/2) + (Technical  Capabilities/3)
+ (Internal  Organization/2) + (Digital Culture)
\end{adjustbox}
\end{equation}

\begin{table}[h!]
\centering
\begin{tabular}{ll}
\hline
\textbf{Indicator}                          & \textbf{Dimension}   \\
\hline
Quality of external links                        & \multirow{4}{*}{Stakeholder Engagement}                 \\
Facebook                           &    \\
Instagram                          &                 \\
Linkedin                           &                 \\
\hline
Best Practice          & \multirow{3}{*}{Technical Capabilities}                \\
Security         &     \\
Speed &                 \\
\hline
URL's length        &         \multirow{2}{*}{Internal Organization}           \\
Quality of internal links   &                 \\      
\hline
Website's age                     & Digital Culture \\
\hline
\end{tabular}
\caption{\label{tab:features}Corporate Digital Assessment Index (CoDAI): indicators and dimensions. }
\end{table}

\begin{table}[h!]
\centering
\small
\begin{adjustbox}{max width=1.\textwidth,center}

\begin{tabular}{llllll}
\toprule
{}&         stakeholder eng. &   technical capabilities &          digital culture &    internal organization &   CoDAI \\
\midrule
Constant &   \makecell{ 0.827*** \\(0.006)} &   \makecell{ 2.346*** \\(0.002)} &   \makecell{ 0.252*** \\(0.003)} &   \makecell{ 0.862*** \\(0.001)} &   \makecell{ 1.672*** \\(0.003)} \\
Micro firms &  \makecell{ -0.076*** \\(0.003)} &  \makecell{ -0.018*** \\(0.001)} &  \makecell{ -0.049*** \\(0.001)} &  \makecell{ -0.003*** \\(0.000)} &  \makecell{ -0.075*** \\(0.002)} \\
Mid-sized firms &   \makecell{ 0.119*** \\(0.005)} &   \makecell{ 0.020*** \\(0.002)} &   \makecell{ 0.040*** \\(0.003)} &   \makecell{ 0.004*** \\(0.000)} &   \makecell{ 0.079*** \\(0.003)} \\
Large firms &   \makecell{ 0.221*** \\(0.010)} &   \makecell{ 0.033*** \\(0.004)} &   \makecell{ 0.058*** \\(0.005)} &   \makecell{ 0.007*** \\(0.001)} &   \makecell{ 0.128*** \\(0.006)} \\
NACE sector A &     \makecell{ 0.029* \\(0.012)} &      \makecell{ 0.009 \\(0.005)} &  \makecell{ -0.028*** \\(0.006)} &     \makecell{ 0.003* \\(0.001)} &    \makecell{ -0.017* \\(0.007)} \\
NACE sector C &   \makecell{ 0.132*** \\(0.004)} &      \makecell{ 0.002 \\(0.002)} &   \makecell{ 0.064*** \\(0.002)} &   \makecell{ 0.015*** \\(0.000)} &   \makecell{ 0.105*** \\(0.003)} \\
NACE sector F &   \makecell{ 0.033*** \\(0.005)} &  \makecell{ -0.014*** \\(0.002)} &  \makecell{ -0.022*** \\(0.003)} &   \makecell{ 0.002*** \\(0.000)} &  \makecell{ -0.018*** \\(0.003)} \\
Nace sector G &   \makecell{ 0.047*** \\(0.004)} &  \makecell{ -0.008*** \\(0.002)} &   \makecell{ 0.037*** \\(0.002)} &   \makecell{ 0.013*** \\(0.000)} &   \makecell{ 0.053*** \\(0.003)} \\
NACE sector H &   \makecell{ 0.039*** \\(0.008)} &  \makecell{ -0.017*** \\(0.003)} &     \makecell{ -0.001 \\(0.004)} &      \makecell{ 0.001 \\(0.001)} &      \makecell{ 0.004 \\(0.005)} \\
NACE sector I &  \makecell{ -0.074*** \\(0.006)} &  \makecell{ -0.011*** \\(0.002)} &   \makecell{ 0.051*** \\(0.003)} &  \makecell{ -0.009*** \\(0.001)} &   \makecell{ 0.024*** \\(0.004)} \\
NACE sector J &   \makecell{ 0.201*** \\(0.006)} &   \makecell{ 0.026*** \\(0.002)} &   \makecell{ 0.079*** \\(0.003)} &   \makecell{ 0.028*** \\(0.001)} &   \makecell{ 0.152*** \\(0.004)} \\
NACE sector K &   \makecell{ 0.149*** \\(0.013)} &   \makecell{ 0.023*** \\(0.005)} &      \makecell{ 0.011 \\(0.007)} &   \makecell{ 0.008*** \\(0.001)} &   \makecell{ 0.060*** \\(0.008)} \\
NACE sector L &   \makecell{ 0.050*** \\(0.009)} &    \makecell{ -0.008* \\(0.004)} &   \makecell{ 0.048*** \\(0.005)} &     \makecell{ -0.000 \\(0.001)} &   \makecell{ 0.058*** \\(0.006)} \\
NACE sector M &   \makecell{ 0.238*** \\(0.006)} &   \makecell{ 0.023*** \\(0.002)} &   \makecell{ 0.048*** \\(0.003)} &   \makecell{ 0.018*** \\(0.001)} &   \makecell{ 0.124*** \\(0.003)} \\
Urban area &   \makecell{ 0.030*** \\(0.003)} &    \makecell{ 0.003** \\(0.001)} &    \makecell{ 0.004** \\(0.001)} &   \makecell{ 0.003*** \\(0.000)} &   \makecell{ 0.014*** \\(0.002)} \\
North &   \makecell{ 0.067*** \\(0.003)} &   \makecell{ 0.020*** \\(0.001)} &   \makecell{ 0.007*** \\(0.002)} &   \makecell{ 0.005*** \\(0.000)} &   \makecell{ 0.033*** \\(0.002)} \\
South &  \makecell{ -0.038*** \\(0.004)} &  \makecell{ -0.024*** \\(0.002)} &  \makecell{ -0.030*** \\(0.002)} &    \makecell{ -0.001* \\(0.000)} &  \makecell{ -0.048*** \\(0.002)} \\
Firm age &  \makecell{ -0.001*** \\(0.000)} &  \makecell{ -0.000*** \\(0.000)} &   \makecell{ 0.007*** \\(0.000)} &  \makecell{ -0.000*** \\(0.000)} &   \makecell{ 0.006*** \\(0.000)} \\
Wide\_band &   \makecell{ 0.018*** \\(0.004)} &     \makecell{ -0.001 \\(0.001)} &     \makecell{ -0.001 \\(0.002)} &      \makecell{ 0.000 \\(0.000)} &      \makecell{ 0.003 \\(0.002)} \\
\midrule
\midrule
R-squared: &                          0.049 &                          0.014 &                           0.17 &                          0.043 &                          0.168 \\
Adj. R-squared: &                          0.048 &                          0.014 &                           0.17 &                          0.043 &                          0.168 \\
\midrule
N. of observations & 182,705 & 182,705 & 182,705 & 182,705 & 182,705 \\
\bottomrule
\end{tabular}
\end{adjustbox}
\caption{\label{tab:allfeatures} regression results considering the CoDAI and its components as dependent variables. For a legend of NACE sectors see table \ref{tab:regression_features}}.

\end{table}

Table \ref{tab:features} shows the dimensions and indicators we used to build the CoDAI. Table \ref{tab:features} summarizes the regression results for the analysis of the determinants of the CoDAI and its four dimensions. Our proposal shows the expected significance level for firm dimension, with the prominence of big firms in comparison to medium and micro. We find interesting results also across sectors. For instance, the case of wholesale and retail
trade (G) shows a positive and significant effect for what concerns stakeholder engagement, but a negative and significant effect for the technical capabilities subpart.
Wideband is significant only for stakeholder engagement. This could be explained by the fact that the majority of the proxies, employed to realize such as category, rely on the adoption of external platforms. 
With no surprise, we find a negative effect for three out of four sub-dimensions of the CoDAI concerning the age of the firm. This could be interpreted as a more reactive digital behavior of newborn digital firms in comparison to more experienced ones. 

Furthermore, the relevance of our proposed framework is evident if we compare the results of our CoDAI with the simple sum of the ten indicators (see table \ref{tab:regression_summed} in the appendix). When we use a simple sum, we can notice how some of the Digital Divide aspects previously pointed out, tend to become blurry and difficult to interpret. For instance, the industrial prominence of Accommodation and food service (I) is in contrast with previous findings on the ICT sector. In addition, the South and North geographical areas signs are inverted and the wideband variable becomes positive and significant.  

\begin{figure}[h!]
    \centering
    \includegraphics[width=\columnwidth]{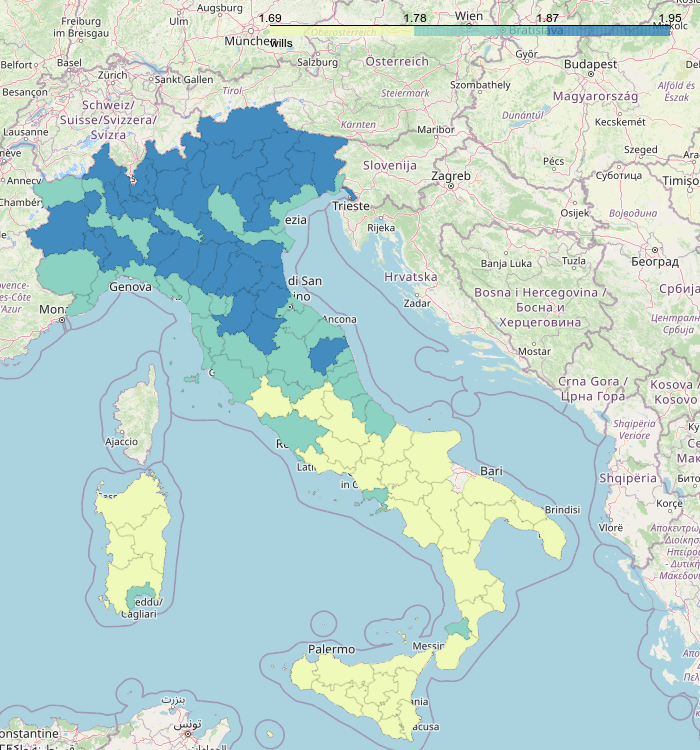} \\
    \caption{The distribution of the CoDAI across Italian regions (NUTS3).}
    \label{fig:wai}
\end{figure}

To highlight the sharp corporate Digital Divide of Italy we map the results of our CoDAI across NUTS regions. Figure \ref{fig:wai} displays three rough geographical clusters, representative of the three main areas of Italy: North, Centre and South (see also figure \ref{fig:wai2} in the appendix for the map at municipal level). This result is noteworthy as confirms the traditional North-South economic divide, using also a novel set of web-related measures.

\section[sec:Discussion and Conclusion]{Final Discussion}\label{sec:conclusion}

Nowadays, the astonishing pervasiveness of ICTs in production and consumption processes has revolutionized how firms operate and strategize. As in every technological transformation, the time and quality of adoption of digital means have created a corporate Digital Divide between more structured and digital-oriented firms and more traditional ones. 
Staying behind this new paradigm has several consequences for the future of small businesses, which characterize the industrial structure of ``digital latecomers'' such as Italy. Moreover, no industry is immune to the increasing birth of hybrid cyber-physical systems which disrupt how products are designed, realized and delivered.

With few exceptions, the extant literature mapped the digital divide phenomenon using data on broadband access, neglecting the role of digital tools and the relative competencies built on it.

With this paper, we continue along the research stream on the use of web scraping to identify digital economy phenomena \citep{Goketal2015,Blazquezetal2018,AxenbeckBreithaupt2019,HerouxVaillancourtetal2020,KinneandAxenback2020,Krugeretal2020,Thoniparaetal2022}. 

In particular, this explorative work introduces a novel way to map and evaluate the corporate digital divide at the granular level, overcoming the country-level perspective. Leveraging the potentiality of web scraping techniques, we inquire about the corporate Digital Divide, extracting, storing and analysing a set of website features related to the digital footprints of 182\,705 Italian firms. We purposefully focus only on technical features as less manipulable in comparison to content and experience and with a less degree of variability.

To improve the comparability of our results across firms, we introduce a Corporate Digital Assessment Index (CoDAI), interpreting the results across four distinctive aspects of the digital strategy of the firm. The CoDAI and its four dimensions confirm the main drivers of the corporate Digital Divide highligted in the literature: we find a prominence of big firms, active in ICT-related fields and localized in more industrialized urban contexts. 

Notwithstanding the relevance of our contribution to the analysis of the corporate Digital Divide, this work is not free of limits.
The literature on the economic interpretation of websites and digital means is still in its infancy. Therefore, the iterative dialogue between empirical testing and theoretical building offers food for thought for multidisciplinary research between economics, management, and data science.

In particular, the relevant elements of a website are changing with an impressive frequency. In this regard, multimedia elements (e.g., images, videos and sounds) have skyrocketed as means of communication thanks to the introduction of new technological enablers (e.g., high-speed broadband), as well as website design has become an important method to establish a website quality \citep{RasmussenandThimm2015}. In addition, also considering the myriad of legal and business services that have progressively emerged thanks to cloud computing, websites are more and more part of the wider ecosystem.

This research opens up new perspectives of data-driven digital monitoring, possibly extending the analysis to more extensive samples of firms across countries, for instance, using the same approach for all firms in ORBIS. The building of proxies able to capture, in economic terms, the digital behaviour of firms represents an essential tool for policymakers. The firm's digital footprint is an important measure to define targeted policies and development strategies for its replicability, unobtrusiveness, frequent updating, extension to new website information, and industrial benchmarking.
In particular, the Italian National Recovery and Resilience Plan (named``NRRP'') established to help the country to recover from the Covid-19 pandemic has 21\%  of the total funds dedicated to digitalisation (with actions such as fastest connection through ultra-broadband, incentives for the adoption of innovative technologies by the private sector, revitalisation of touristic and cultural sectors) \footnote{see https://www.governo.it/it/approfondimento/le-missioni-e-le-componenti-del-pnrr/16700.}.
The building of new tools and methodologies, such as the ones we provide with this work, able to identify laggard territories and NUTS-3 regions, represents a crucial step in an era in which policies are more and more data-driven. With the awareness of the limit and potentiality of our analysis, we contribute to this research stream with an innovative and original approach, which joint public-private strategic partnerships can further leverage.

\pagebreak
\bibliographystyle{tfcad}
\bibliography{ref}
\pagebreak
\section{Appendix}

\begin{table}[ht!]
\centering
\small
\begin{adjustbox}{max width=1.\textwidth,center}
\begin{tabular}{ll}
\toprule
{}&                     sum of the ten indicators \\
\midrule
Constant &   \makecell{ 2.601*** \\(0.011)} \\
Micro firm &  \makecell{ -0.178*** \\(0.006)} \\
Mid-sized firm &   \makecell{ 0.244*** \\(0.010)} \\
Large firm&   \makecell{ 0.416*** \\(0.019)} \\
A &   \makecell{ 0.115*** \\(0.023)} \\
C &  \makecell{ -0.064*** \\(0.008)} \\
G &   \makecell{ 0.088*** \\(0.008)} \\
F &  \makecell{ -0.299*** \\(0.011)} \\
H &  \makecell{ -0.321*** \\(0.015)} \\
I &   \makecell{ 0.311*** \\(0.011)} \\
J &   \makecell{ 0.172*** \\(0.011)} \\
M &   \makecell{ 0.071*** \\(0.011)} \\
K &  \makecell{ -0.205*** \\(0.024)} \\
L &   \makecell{ 0.101*** \\(0.018)} \\
Urban area &   \makecell{ 0.055*** \\(0.005)} \\
North &  \makecell{ -0.026*** \\(0.006)} \\
South &     \makecell{ 0.019* \\(0.008)} \\
Firm age &   \makecell{ 0.001*** \\(0.000)} \\
Wide band &   \makecell{ 0.027*** \\(0.007)} \\
\midrule
\midrule
R-squared: &                          0.038 \\
Adj. R-squared: &                          0.038 \\
\midrule
N. of Observations & 182705  \\

\bottomrule
\end{tabular}
\end{adjustbox}
\caption{\label{tab:regression_summed} OLS regression results. The dependent variable is the sum of the values of the ten indicators of the corporate web sites.}
\end{table}

\begin{figure}[h!]
    \centering
 \includegraphics[width=2\columnwidth]{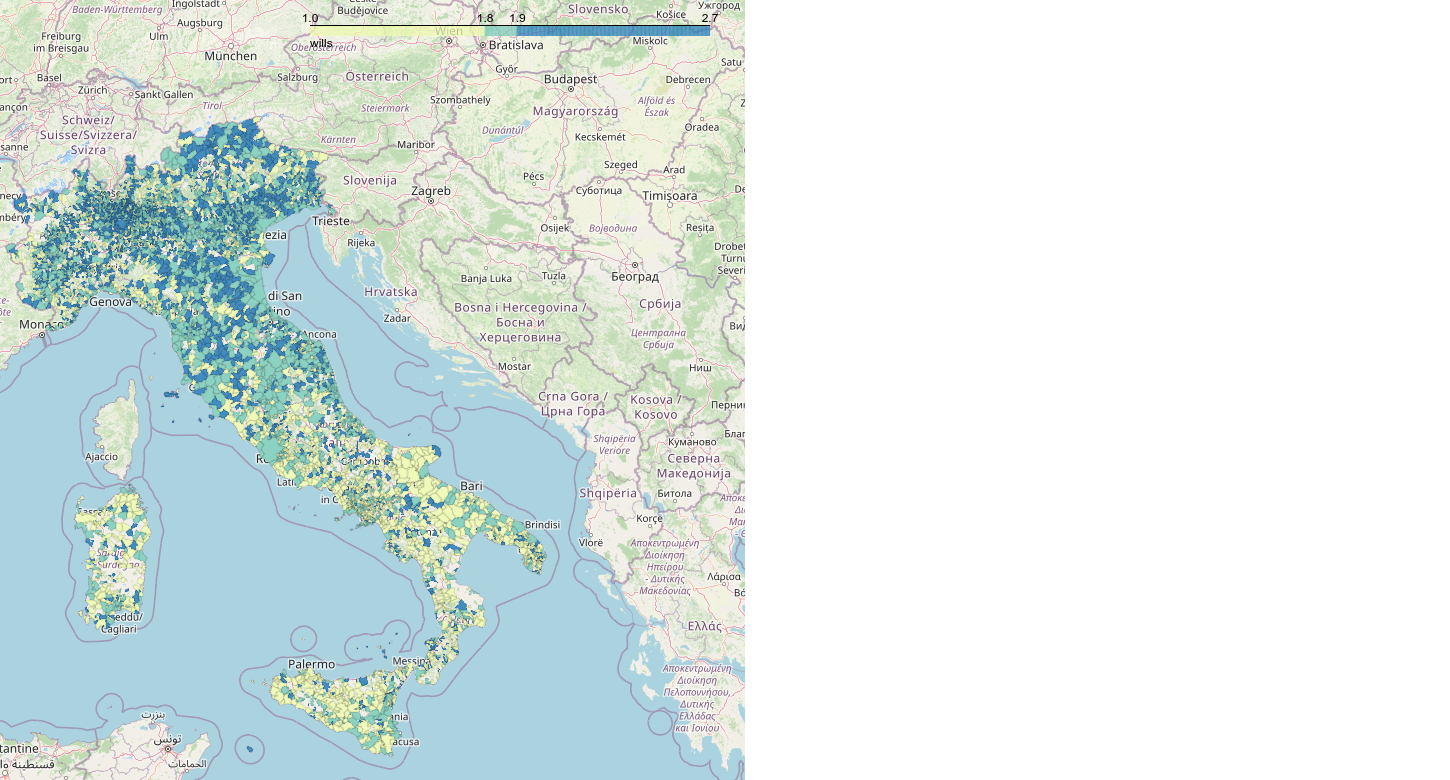}\\
    \caption{The distribution of the CoDAI across Italian municipalities.}
    \label{fig:wai2}
\end{figure}

\end{document}